\documentclass[reprint,aps,prl,superscriptaddress]{revtex4-2}
\setcitestyle{super}
\usepackage[version=4]{mhchem}
\usepackage{amsmath}
\usepackage{amsfonts}
\usepackage{amssymb}
\usepackage[noabbrev]{cleveref}
\usepackage{bm}
\usepackage{txfonts}
\usepackage{physics}
\usepackage{lipsum}
\usepackage{graphicx}
\usepackage[normalem]{ulem}
\usepackage{here}
\usepackage{multirow}
\usepackage{threeparttable}
\usepackage[usenames]{xcolor}

\def\e{\varepsilon }

\def\AA{$\mathrm{\mathring{A}}$}

\def\tilde{\widetilde}

\allowdisplaybreaks

\begin{document}

\title{Oxirane Ring-Opening Dynamics Mediated By The First Singlet Excitation Using Time-Dependent \textit{GW} Molecular Dynamics}

\author{Aaditya Manjanath}
\email[Corresponding author: ]{manjanath.aaditya@nims.go.jp}
\affiliation{Research Center for Structural Materials, National Institute for Materials Science, 1-2-1 Sengen, Tsukuba 305-0047, Ibaraki, Japan}
\author{Ryoji Sahara}
\affiliation{Research Center for Structural Materials, National Institute for Materials Science, 1-2-1 Sengen, Tsukuba 305-0047, Ibaraki, Japan}
\author{Thi Nu Pham}
\affiliation{Department of Physics, Graduate School of Engineering, Yokohama National University, 79-5 Tokiwadai, Hodogaya-ku, Yokohama 240-8501, Kanagawa, Japan}
\author{Yoshiyuki Kawazoe}
\affiliation{New Industry Creation Hatchery Center, Tohoku University, 6-6-4 Aramaki Aza Aoba, Aoba-ku, Sendai 980-8579, Miyagi, Japan}
\affiliation{Centre for Interdisciplinary Research, SRM University-AP, Amaravati
522240, Andhra Pradesh, India}
\author{Kaoru Ohno}
\email[Corresponding author: ]{ohno@ynu.ac.jp}
\affiliation{Department of Physics, Graduate School of Engineering, Yokohama National University, 79-5 Tokiwadai, Hodogaya-ku, Yokohama 240-8501, Kanagawa, Japan}

\begin{abstract}
Oxirane (ethylene oxide, \ce{C2H4O}) is a prototypical system for investigating photoinduced bond cleavage and ring-opening dynamics. Previous time-dependent density functional theory studies identified the initial HOMO $\rightarrow$ LUMO (\ce{S1}) excitation as nonreactive and attributed ring opening to higher-energy excitations. Subsequent constrained density functional theory simulations, however, reported ring opening following an \ce{S2} $\rightarrow$ \ce{S1} transition. Here, we investigate photochemistry of oxirane using time-dependent $GW$ molecular dynamics (TD$GW$-MD), a many-body approach based on extended quasiparticle theory [\textit{J. Chem. Phys.} \textbf{160}, 184102 (2024)]. Our simulations reveal ring opening initiated by asymmetric \ce{C-O} bond cleavage following direct HOMO $\rightarrow$ LUMO (\ce{S1}) excitation. Following the selected surface hop, separate conditional trajectories on the broken-symmetry open-shell singlet and triplet potential energy surfaces lead to distinct products. The singlet trajectory yields acetaldehyde via intramolecular hydrogen transfer, while the triplet trajectory produces formaldehyde and methylene via \ce{C-C} bond fragmentation. These products are consistent with those identified in previous photolysis experiments. Our results demonstrate that direct excitation to the lowest singlet excited state can initiate ring opening within the present TD$GW$-MD framework. They further reveal distinct subsequent reaction pathways under the selected spin conditions. This study highlights the potential of TD$GW$-MD for elucidating photoinduced reaction mechanisms from a many-body perspective.
\end{abstract}


\maketitle




\section{Introduction}

Oxirane (ethylene oxide, \ce{C2H4O}) is a prototypical system for investigating photoinduced ring opening and the reactions that follow bond cleavage \cite{GomerNoyes1950, Roquitte1966, Takezaki1973, Cordova2007, Tapavicza2008, HuixRotllant2010, Li2015, Jenkins2019, Schmidt2020}. Its ultraviolet photolysis produces several rearrangement and dissociation products, as established by the early experiments of Gomer and Noyes and subsequent vacuum-ultraviolet studies \cite{GomerNoyes1950, Roquitte1966, Takezaki1973}. In the Gomer-Noyes pathway, asymmetric \ce{C-O} cleavage opens the three-membered ring, followed by intramolecular hydrogen transfer to form acetaldehyde and, in some cases, further fragmentation into methyl (\ce{CH3^{.}}) and formyl (\ce{HCO^{.}}) radicals \cite{GomerNoyes1950}. Beyond this mechanism, depending on the excitation condition, other dissociation products such as \ce{O + C2H4} (\ce{C2H2 + H2}), \ce{H2 + CH2CO}, and \ce{CH2 + HCHO} have been reported through experiments \cite{Takezaki1973}. The ring-opening reaction involves substantial changes in orbital character and the formation of an open-shell intermediate, making oxirane a suitable target for investigating its rich photochemistry using established or newly developed excited-state electronic-structure methods \cite{Cordova2007, Tapavicza2008, HuixRotllant2010, Li2015}. Its small size also permits comparison with multiconfigurational and quantum Monte Carlo (QMC) approaches, which have been used to assess the potential-energy surfaces sampled during the reaction \cite{Cordova2007, Tapavicza2008, HuixRotllant2010}. \\

Previous simulations have reproduced the Gomer-Noyes mechanism reasonably well but differ in their assignment of the excitation for initiating the ring-opening process. Using time-dependent density functional theory (TDDFT) \cite{Runge1984, Petersilka1996} within the Tamm–Dancoff approximation (TDA), i.e., TDDFT/TDA, and classical trajectory surface hopping (TSH) \cite{Tully1971, Tully1990}, Tapavicza \textit{et al.} identified the initial ($n\rightarrow3s$), HOMO$\rightarrow$LUMO (\ce{S1}) channel as nonreactive and associated ring-opening with the higher ($n\rightarrow3p_z$), \ce{S2} channel, which is an apparent departure from Kasha's rule \cite{Kasha1950, Tapavicza2008}. They reported that ring-opening occurred either after direct excitation to \ce{S2} or when the two singlet excited states \ce{S1} and \ce{S2} mix for an \ce{S1}-initialized trajectory \cite{Tapavicza2008}. Using quantified natural transition orbital analysis along a constrained ring-opening coordinate, Li \textit{et al.} identified a downhill pathway associated with $n\rightarrow3p_z$ character, whereas the $n\rightarrow3s$ pathway encountered a barrier before reaching a crossing with the $n\rightarrow3p_z$ reactive channel \cite{Li2015}. In contrast, Krenz \textit{et al.} found ring opening to be quantitatively correlated with populating the \ce{S1} state following an \ce{S2}$\rightarrow$ \ce{S1} hop in occupation-constrained density functional theory (DFT) simulations initialized in the \ce{S2} state \cite{Schmidt2020}. Bin \textit{et al.} also identified ring-opening pathways using ensemble DFT with a range-separated hybrid functional \cite{Jenkins2019}. However, their lowest excited state (ES) was assigned $n \rightarrow \sigma^*$ character and placed substantially above the observed lowest excitation \cite{Tapavicza2008, Jenkins2019, Schmidt2020}, distinguishing their pathways from those initiated in the low-lying $n \rightarrow 3s$ Rydberg state. Together, these studies make the description of the initially excited state central to the interpretation of oxirane ring opening. Assessing the proposed mechanisms therefore requires comparison of the low-lying excitation spectrum with spectroscopy, together with examination of the dynamics following direct preparation of the lowest excitation. \\


The preceding dynamics studies employed density functional approaches that in turn rely on the adiabatic approximation (AA). However, AA neglects density-history and explicit initial-state dependence in the exchange-correlation (XC) response \cite{Cordova2007, HuixRotllant2010, Elliott2012}, limiting its reliability for dynamics initiated from electronically ESs \cite{Lacombe2023}. These limitations motivate an alternative description of electronic excitations. \\

Quasiparticle (QP) theory based on many-body perturbation theory (MBPT) treats electron-electron interactions through the self-energy \cite{Hedin1965, Csanak1971, Strinati1980, Strinati1982, Strinati1988}. Within the $GW$ approximation \cite{Strinati1980, Strinati1982, Hybertsen1985, Hybertsen1986}, this nonlocal, energy-dependent self-energy accounts for exchange and dynamically screened electron correlation. This approach has substantially corrected the band-gap underestimation of common density-functional approximations in semiconductors and insulators \cite{Hybertsen1986}. This success motivates investigating a corresponding QP description of dynamics initiated from electronically ESs, with the aim of examining photoinduced bond cleavage. \\

Ohno {\it et al.} showed that the QP approaches within MBPT such as the one-shot $GW$ approximation \cite{Hybertsen1986, Ishii2003, Aoki2018} using the local spin-density approximation \cite{Perdew1981} to generate reference eigeninformation, self-consistent $GW\Gamma$ \cite{Kuwahara2016}, and $GW +$ Bethe--Salpeter equation \cite{Rohlfing2000, Noguchi2010, Ohno2024} can directly be applied to any excited eigenstate as the reference state \cite{Isobe2018, Aoki2019, Ohno2022} thanks to extended quasiparticle theory (EQPT) \cite{Ohno2017, Ohno2018Book}. In EQPT, electron-addition and electron-removal energies are defined for any arbitrary initial excited (or ground) eigenstate of the many-body Hamiltonian \cite{Ohno2017, Ohno2018Book}. It also satisfies extended Koopmans' theorem \cite{Morrell1975, Smith1975}. A time-dependent extension of EQPT can therefore describe dynamics initiated from an electronically excited eigenstate without invoking the adiabatic XC approximation of TDDFT, thereby circumventing the associated neglect of explicit initial-state dependence. \\

We recently developed non-adiabatic excited-state time-dependent $GW$ molecular dynamics (NA-ES-TD$GW$-MD or TD$GW$-MD for short) \cite{Manjanath2024a, Manjanath2024b}, based on EQPT, in which the time-dependent QP equation was straightforwardly derived from the time-dependent many-body Schr\"{o}dinger equation. In our practical implementation \cite{Manjanath2024a, Manjanath2024b}, one-shot $GW$ corrects the Kohn-Sham (KS) eigenvalues obtained from the usual DFT procedure while approximating QP wavefunctions by KS orbitals \cite{Hybertsen1986, Ishii2003, Aoki2018}, and the local density approximation (LDA) energy gradient provides the nuclear forces with limited QP contributions to the total energy. It has been successfully applied to methane and nickel-assisted methane photolysis \cite{Manjanath2024a, Manjanath2024b}, building on earlier work on successive hydrogenation of carbon monoxide \cite{Pham2016}. \\

In this work, we apply TD$GW$-MD to oxirane following direct HOMO $\rightarrow$ LUMO excitation, associated with the initial ($n\rightarrow3s$) transition identified as the \ce{S1} state \cite{Cordova2007, Tapavicza2008, Schmidt2020}. We first compare the calculated ionization and low-lying excitation energies with spectroscopic data, then examine the dynamics following direct preparation of this excitation, which produces asymmetric \ce{C-O} cleavage and ring opening. This demonstrates that the lowest excitation is sufficient to initiate this reaction in the present simulation. We subsequently perform surface hops to two different branches (singlet and triplet) from the ring-opened geometry. The broken-symmetry open-shell singlet trajectory yields acetaldehyde through intramolecular hydrogen transfer, whereas the triplet trajectory produces formaldehyde and methylene through \ce{C-C} bond fragmentation. These products are consistent with those observed in previous photolysis experiments \cite{GomerNoyes1950, Takezaki1973}. \\

These conditional trajectories establish the different chemical pathways accessible on the selected spin surfaces, without determining the product yields. These results demonstrate that a many-body approach such as TD$GW$-MD can be extended to such non-trivial ring-opening processes and can correctly identify electronic rearrangements accompanying bond cleavage and product formation.

\section{Methods}
\label{Method}

\subsection{Time-dependent QP equations}

Similar to the time-dependent Kohn-Sham equation \cite{Runge1984},
the time-dependent QP equation (atomic units are used throughout unless otherwise stated)
\begin{align}
{\rm i}\frac{\partial}{\partial t}\phi_{\lambda}(\mathbf{r};\mathbf{R}(t),t)
= {\cal H}^{\rm QP}_{\rm mixed}(\mathbf{r};\mathbf{R}(t))\phi_{\lambda}(\mathbf{r};\mathbf{R}(t),t),
\label{TDQPeq}
\end{align}
holds for the wavepackets $\phi_{\lambda}(\mathbf{r};\mathbf{R}(t),t)$,
where $\mathbf{R}(t)=\{\mathbf{R}_I(t)\}$ represents a set of nuclear coordinates ($I$ - atomic index).
Here, ${\cal H}^{\rm QP}_{\rm mixed}(\mathbf{r};\mathbf{R}(t))$ stands for
the QP ($GW$) Hamiltonian for the mixed state \cite{Manjanath2024a}.
The static counterpart of Eq. (\ref{TDQPeq}) reads 
\begin{align}
{\cal H}^{\rm QP}_{\rm mixed}(\mathbf{r};\mathbf{R}(t))\varphi^{\rm QP}_n(\mathbf{r};\mathbf{R}(t))
=\e^{\rm QP}_n(\mathbf{R}(t))\varphi^{\rm QP}_n(\mathbf{r};\mathbf{R}(t)),
\label{eigeneq}
\end{align}
where $\varphi^{\rm QP}_n(\mathbf{r};\mathbf{R}(t))$ and $\e^{\rm QP}_n(\mathbf{R}(t))$
denote, respectively, the QP wavefunctions and the QP energies [$n=i$ (occupied) or $a$ (empty)].
Using the spectral method \cite{Sahara2013,Pham2016,Manjanath2024a,Manjanath2024b},
we can expand the wavepackets $\phi_{\lambda}(\mathbf{r};\mathbf{R}(t+\Delta t),t+\Delta t)$
at a slightly later time as
\begin{widetext}
\begin{align}
\phi_{\lambda}(\mathbf{r};\mathbf{R}(t+\Delta t),t+\Delta t)
= {\rm exp}\left[-{\rm i}\int_t^{t+\Delta t}{\cal H}^{\rm QP}_{\rm mixed}(\mathbf{r};\mathbf{R}(t))dt\right]\,
\phi_{\lambda}(\mathbf{r};\mathbf{R}(t),t)
\approx \sum_n {\rm exp}\left[ -{\rm i}\e^{\rm QP}_n(\mathbf{R}(t+\Delta t))\Delta t \right]\,
\varphi^{\rm QP}_n(\mathbf{r};\mathbf{R}(t+\Delta t)) c_{n\lambda}(t+\Delta t)
\label{spectral2}
\end{align}
\end{widetext}
with
\begin{widetext}
\begin{align}
c_{n\lambda}(t+\Delta t)
&= \langle\,\varphi^{\rm QP}_n(\mathbf{R}(t+\Delta t))\,|\,\phi_{\lambda}(\mathbf{R}(t),t)\,\rangle
\approx \langle\,\varphi^{\rm QP}_n(\mathbf{R}(t))\,|\,\phi_{\lambda}(\mathbf{R}(t),t)\,\rangle
+ \Delta t \left[\frac{\partial}{\partial t'}
\langle\,\varphi^{\rm QP}_n(\mathbf{R}(t'))\,|\,\phi_{\lambda}(\mathbf{R}(t),t)\,\rangle\right]_{t'=t}
\nonumber\\
&= c_{n\lambda}(t) - \Delta t \sum_m c_{m\lambda}(t)
\sum_I\dot{\mathbf{R}}_I(t)
\bigl\langle\,
\varphi^{\rm QP}_n(\mathbf{R}(t))\,\big|\,\nabla_{\mathbf{R}_I}\,\big|\,\varphi^{\rm QP}_m(\mathbf{R}(t))
\,\bigr\rangle.
\label{coeff_dt}
\end{align}
\end{widetext}
The second term represents a non-adiabatic interaction \cite{Saalmann1996,Makhov2014},
where $\dot{\mathbf{R}}_I(t)=d\mathbf{R}_I(t)/dt$ denotes the nuclear velocity.
The matrix elements $\mathbf{d}_{nm}^I = \bigl\langle\,
\varphi^{\rm QP}_n(\mathbf{R}(t))\,\big|\,\nabla_{\mathbf{R}_I}\,\big|\,\varphi^{\rm QP}_m(\mathbf{R}(t))\,\bigr\rangle$ represent the non-adiabatic coupling vectors. 

\subsection{One-shot \textit{GW} within TDQP}

We use the one-shot $GW$ approach \cite{Hybertsen1986}, where the QP wavefunctions are replaced by
Kohn-Sham orbitals obtained by using the LDA, i.e., $\varphi^{\rm QP}_n(\mathbf{r};\mathbf{R}(t)) \approx \varphi^{\rm LDA}_n(\mathbf{r};\mathbf{R}(t))$, and the QP energy eigenvalues $\varepsilon^{\rm QP}_n(\mathbf{R}(t))$ are non--self-consistent corrections to the LDA eigenvalues $\varepsilon^{\rm LDA}_n(\mathbf{R}(t))$,
\begin{widetext}
\begin{align}
\e^{\rm QP}_n(\mathbf{R}(t)) \approx \e^{\rm LDA}_n(\mathbf{R}(t))
+ \bigl\langle\varphi^{\rm LDA}_n(\mathbf{R}(t))
\bigm|[\Sigma_{\rm xc}(\e^{\rm QP}_n(\mathbf{R}(t)))-\mu^{\rm LDA}_{\rm xc}]\bigm|
\varphi^{\rm LDA}_n(\mathbf{R}(t))\bigr\rangle,
\label{GW_correction}
\end{align}
\end{widetext}
where $\mu^{\rm LDA}_{\rm xc}(\mathbf{r})$ and $\Sigma_{\rm xc}(\e^{\rm QP}_n(\mathbf{R}(t)))$ denote the XC potential from KS-DFT and the self-energy, respectively. $\Sigma_{\rm xc}(\e^{\rm QP}_n(\mathbf{R}(t)))$ is calculated using the generalized plasmon-pole (GPP) approximation \cite{Hybertsen1986}.
Since the QP energies $\e^{\rm QP}_n$ appearing in the argument of the self-energy $\Sigma_{\rm xc}(\e^{\rm QP}_n(\mathbf{R}(t)))$ can be replaced by the QP energy obtained in the previous time step, the renormalization procedure using the $Z$ factor \cite{Hybertsen1986}
is not required except when $\e^{\rm QP}_n$ is computed for the first time.
Using the QP energies $\e^{\rm QP}_n(\mathbf{R}(t))$,
obtained from Eq.~(\ref{GW_correction}) in the propagation expression, Eq.~(\ref{spectral2}),
is the distinguishing feature of the TD$GW$ approach.
Besides this, everything else remains the same as in the TDDFT dynamics formulation. \\

Owing to the mixed quantum-classical nature of the dynamics approach, the nuclei are treated classically within the Newtonian framework where the atomic forces are computed as
\begin{align}
M_I\frac{d^2\mathbf{R}_I(t)}{dt^2}
\approx -\nabla_{\mathbf{R}_I}E^{(N)}_{\rm LDA}(\mathbf{R}(t),t),
\label{Newton}
\end{align}
where $E^{(N)}_{\rm LDA}(\mathbf{R}(t),t)$ denotes the LDA total energy of an $N$-electron reference state.
We used a variational force for the derivative of the total energy, as formulated by Ho {\it et al.} \cite{Ho1992}. \\

The usage of the LDA total energy in this force formulation to compute atomic forces is justified as the change in the LDA and $GW$ total energies [$GW$ total energy obtained using self-consistent $GW$ \cite{Kuwahara2014}] with the atomic displacement (i.e., the slope of an energy vs. atomic displacement plot) is nearly the same for small atomic displacements \cite{Manjanath2024a}. The true force for dynamics may lie between these two extremes, which is currently indeterminable. Therefore, if ring opening is observed in oxirane using the trajectory obtained with the LDA total energy, it is expected that this is possible even when the true force is computed. 
The true force may be somewhere between the LDA force and the $GW$A force.
The present on-the-fly simulation should therefore be regarded as providing a nearly accurate description of the dynamics along this approximate trajectory.
Besides this subtle difference, the present TD$GW$-MD approach on the basis of the one-shot $GW$ approximation has distinct advantages over the conventional TDDFT-MD approach, for performing excited-state dynamics simulations.

\subsection{\textit{Ab initio} cloning in TD\textit{GW}-MD}

In our approach, the nuclei evolve under the mean-field forces generated by the electrons, which is fundamentally Ehrenfest dynamics. In this mean-field approximation, the correlation between the electron ``motion" and nuclear trajectory is neglected, limiting the description of electronic decoherence. The mean-field description suffers from the problem of strong non-adiabatic mixing particularly when two Born–Oppenheimer (BO) surfaces cross or approach each other. A proper description of such correlations requires a distinct classical trajectory for each electronic state, which is provided by a surface-hopping (SH) strategy \cite{Tully1971, Tully1990, Makhov2014}. We use the SH approach proposed by Makhov \textit{et al.} \cite{Makhov2014} named {\it ab initio} (multiple) cloning in our dynamics formalism, because it is compatible with our single-particle-level approach. Unlike the strategy employed by Makhov \textit{et al.}, however, our approach does not include multiple trajectory basis functions, for simplicity. \\

A diagnostic for leaving the mean-field description, i.e., hopping from a mixed surface (with index $\lambda$) to a pure BO surface (with index $n$) is called the ``breaking force" $\mathbf{F}^{\rm br}_{\lambda \rightarrow n}$. The indices $\lambda$ and $n$ typically indicate the newly occupied/empty QP level (OCC1/EMP1, see Section \ref{Results}) in the QP representation. The breaking force is defined as
\begin{align}
\mathbf{F}^{\rm br}_{\lambda \rightarrow n} = (1 - |c_{n\lambda}(t)|^2)\Delta\mathbf{F}_{n\lambda}(t),
\end{align}
where $c_{n\lambda}(t)$ are the coefficients in the expansion of the wavepacket as defined in Eq.~(\ref{coeff_dt}) and $\Delta\mathbf{F}_{n\lambda}(t)$ is the deviation of the force felt by the nuclei on a pure BO surface ($-\nabla_{\mathbf{R}_I}E^{(N)}_n(\mathbf{R}(t),t)$) from the mean-field force in Eq.~(\ref{Newton}) [$-\nabla_{\mathbf{R}_I}E^{(N)}_{\lambda}(\mathbf{R}(t),t)$]:
\begin{align}
\Delta\mathbf{F}_{n\lambda}(t) = \nabla_{\mathbf{R}_I}E^{(N)}_{\lambda}(\mathbf{R}(t),t)
- \nabla_{\mathbf{R}_I}E^{(N)}_n(\mathbf{R}(t),t).
\end{align}
$\mathbf{F}^{\rm br}_{\lambda \rightarrow n}$ is mass-scaled to obtain the so-called \textit{breaking acceleration} $a^{\rm br}_{\lambda\rightarrow n}$ [Eq. (\ref{clone_condition})], which is used to decide when to perform the hop \cite{Makhov2014}. A possible hopping point is identified when $a^{\rm br}_{\lambda\rightarrow n}$ exceeds a predefined threshold $\delta_{\rm clone}$,
\begin{align}
a^{\rm br}_{\lambda\rightarrow n} = |\mathbf{M}^{-1}\mathbf{F}^{\rm br}_{\lambda\rightarrow n}|
> \delta_{\rm clone}.
\label{clone_condition}
\end{align}
Based on our previous experience \cite{Manjanath2024a, Manjanath2024b}, we have found $\delta_{\rm clone} = 3 \times 10^{-6}$ a.u. to be sufficient for deciding on surface hops.

\subsection{Velocity rescaling for post-hop dynamics}

At the hopping geometry, the nuclear velocities were adjusted to conserve the total energy using the momentum-direction prescription discussed and benchmarked by Barbatti \cite{Barbatti2021}. Because the state-to-state nonadiabatic-coupling vector is not available within the present implementation, the velocity adjustment was performed along the full nuclear linear-momentum vector. For each nucleus $A$, the post-hop velocity is given by
\begin{equation}
    \mathbf{v}_{A}^{\mathrm{post}}
    =
    \mathbf{v}_{A}^{\mathrm{pre}}
    +
    \gamma\frac{\mathbf{p}_{A}^{\mathrm{pre}}}{M_A}
    =
    (1+\gamma)\mathbf{v}_{A}^{\mathrm{pre}},
\end{equation}
where $\mathbf{p}_{A}^{\mathrm{pre}}=M_A\mathbf{v}_{A}^{\mathrm{pre}}$.
The parameter $\gamma$ is chosen such that
\begin{equation}
    K_{\mathrm{post}}-K_{\mathrm{pre}}
    =
    -\left(E_{\mathrm{post}}-E_{\mathrm{pre}}\right),
\end{equation}
where $E_{\mathrm{post}}-E_{\mathrm{pre}}$ denotes the electronic energy change associated with the selected hop, estimated from the QP energies using a cationic reference at the hopping geometry. For the downward hops considered here, the decrease in electronic energy is transferred to the nuclear kinetic energy. The solution with the smaller absolute value of $\gamma$ is selected to minimize the perturbation introduced by the velocity adjustment. Consequently, all atomic velocity vectors retain their directions and are multiplied by a common factor. This prescription
was applied to all the trajectories explored in this work.

\subsection{Computational workflow}

We used the all-electron mixed-basis approach \cite{Kuwahara2016,Manjanath2024a,Manjanath2024b,Pham2016,Kuwahara2014,Sahara2013,Ohno2018,Ohtsuki2021} implemented in our home-grown \textit{ab initio} package named Tohoku mixed basis orbitals (TOMBO) \cite{Ono2015}, in which the one-electron orbitals are expressed by both plane waves (PWs) and atomic orbitals (AOs). The isolated oxirane molecule was optimized using LDA in TOMBO. The details of the optimized structure are provided in Section \ref{Results}. \\

We used simple cubic unit cells of lengths 15~\AA~and 12~\AA~for the photoabsorption energy calculations and TD$GW$-MD simulations, respectively. The Coulomb interaction at these lengths was spherically cut to avoid interactions with adjacent
unit cells. The 28671 (14147 for the 12~\AA~cell) PWs corresponding to the 17.7 Ry (17.3 Ry for 12~\AA) cutoff energy were used. A minimal number
of numerical AOs were constructed and have finite values only within each non-overlapping atomic sphere.
AOs were smoothly truncated by subtracting a smooth quadratic function, which has the
same amplitude and derivative at the atomic sphere surface \cite{Ono2015}. This quadratic function
smoothly connecting to the tail of the true orbital can be successfully described by a
linear combination of PWs. The cutoff energies corresponding to 70.9 Ry (69.1 Ry for 12~\AA) and 11.1 Ry were set
for the exchange and correlation parts, respectively, of the self-energy. A total of 120 occupied and empty levels were used
in the spectral decomposition \cite{Manjanath2024a}; 108 empty levels were used for the calculation of the correlation part of the self-energy. The generalized plasmon pole model \cite{Hybertsen1986} was used to avoid frequency integration.
We performed tests to ensure that all these parameters were sufficient for obtaining good
convergence of results. However, we increased the number of levels to 4000 when
separate one-shot $GW$ calculations were performed. \\

The precursor to performing a TD$GW$-MD simulation is to obtain a
converged electronic state at a given excited electron configuration within the LDA. 
In our work, we used a level selection algorithm based on a version of the initial maximum overlap  method \cite{Barca2018} applied only to the frontier orbitals participating in the electron excitation, to ensure self-consistent field (SCF) convergence at the ES.
Once convergence was achieved, the LDA eigenvalues were corrected by the one-shot $GW$ approximation
by virtue of EQPT. Subsequently, the TD$GW$-MD simulation commenced by 
updating the wavepackets stepwise in $\Delta t = 0.01$ fs intervals,
where the Hamiltonian is expected to change only very slightly.
The wavepackets are the same as the QP (KS) eigenfunctions for the ES
reference at $t = 0$. 
Following each update of the nuclear positions, we performed 20 sub-loops of updates to the electronic structure at fixed nuclear coordinates to improve numerical conservation of the total energy.
(All dynamics simulations were performed in the NVE ensemble.)
The $GW$ calculation was performed only at the final (20th) sub-loop, and
the QP energies and atomic positions were updated subsequently. To accelerate the simulations, we performed this one-shot $GW$ calculation every 10 MD steps.


\section{Results and Discussion}
\label{Results}

\subsection{Structural information}

\begin{figure}[!h]
\begin{center}
\includegraphics[width=\columnwidth]{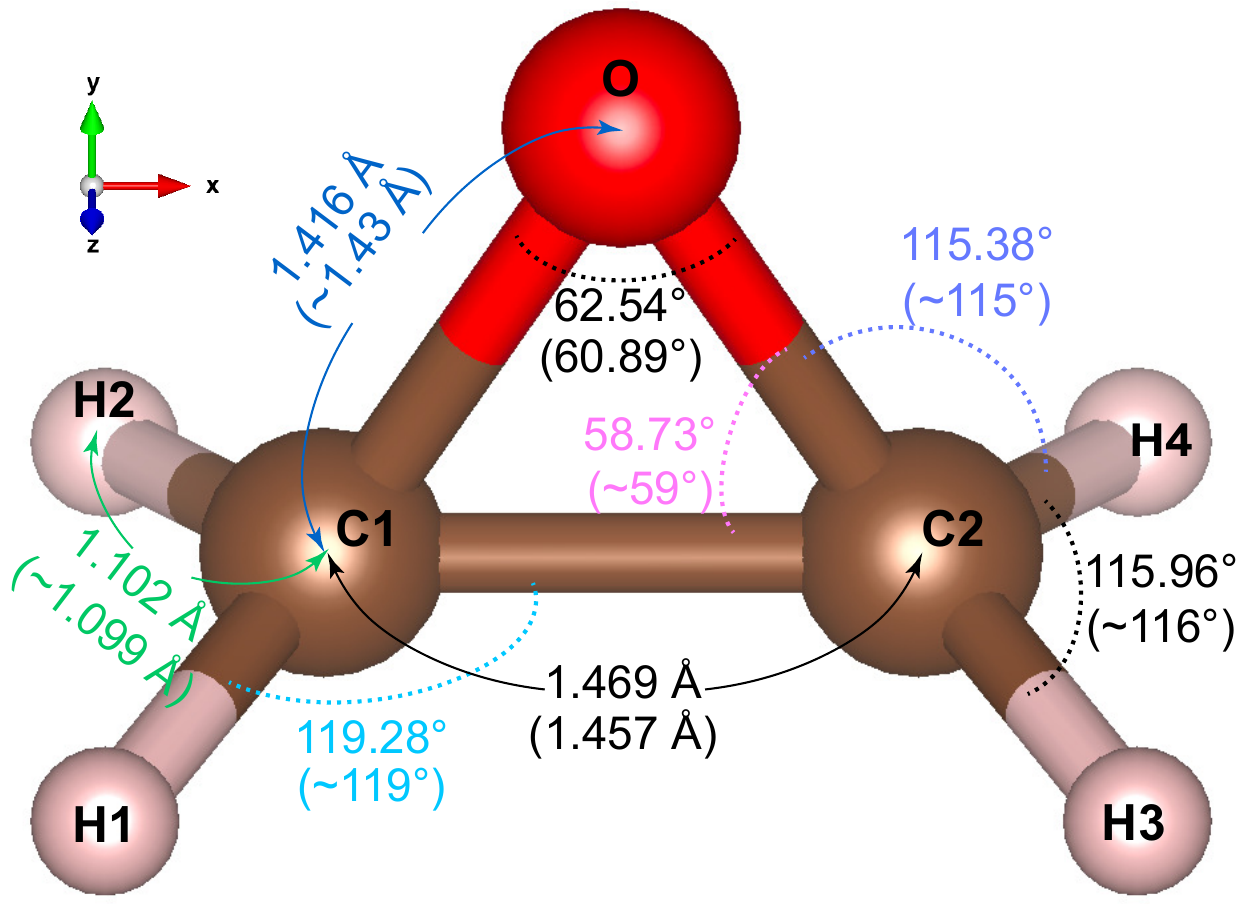}
\caption{Optimized geometry of oxirane. The bond angles are indicated by dotted lines and the bond lengths, by solid lines. Different colors are used to improve readability of the bond lengths and bond angles. In this \textit{C\textsubscript{2v}} symmetric structure, the four C–H bond lengths are equal, as are the two C–O bond lengths. Bond angles related by symmetry are also equal; therefore, only one value is shown for each set of equivalent bond lengths or angles. The values denoted in parentheses are experimental parameters sourced from Refs \citenum{Hirose2006, Grabowsky2008, NIST}. The viewing orientation is indicated.}
\label{F-1}
\end{center}
\end{figure}

We first optimized the initial structure of an isolated oxirane molecule in the electronic ground state (GS) in the gas phase.
The optimized geometry is depicted in Fig. \ref{F-1} with experimental structural parameters from Refs. \citenum{Hirose2006, Grabowsky2008, NIST} shown in parentheses.
There is good agreement between the structural data obtained by LDA in TOMBO and those from experimental reports demonstrating that the LDA level of theory is sufficient for GS structural optimizations.

\subsection{Ground state electronic structure of oxirane}

\begin{figure*}[!ht]
\begin{center}
\includegraphics[width=0.8\textwidth]{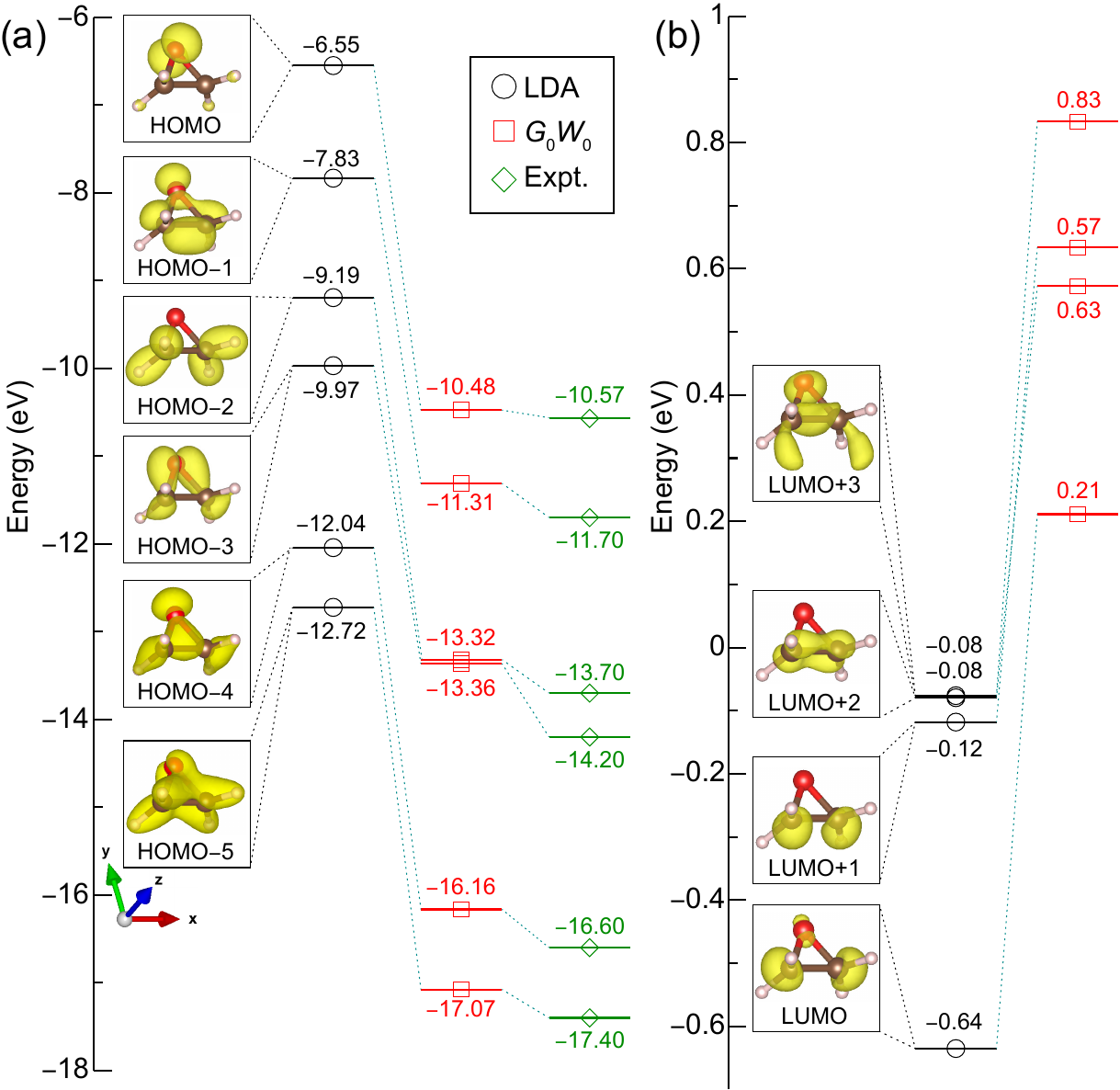}
\caption{Electronic structure of oxirane showing the energy levels obtained from LDA (black circles), \ce{\textit{G}0\textit{W}0} (red squares), and experiments (green diamonds) for the (a) occupied and (b) virtual subspaces. The LDA electron wavefunction-squared (charge density) for each level is also shown enclosed by a box of the same color (black) as the circles representing the LDA energies. The numerical value above each level signifies the energy in eV of that level. The dotted lines are guides to the eye for easy identification of levels from the computed/experimental results (turquoise) or wavefunction plots (black). The viewing orientation is indicated. The isosurface values for the occupied and unoccupied wavefunction squared plots are $5 \times 10^{-9}$ and $5 \times 10^{-10}$ electrons/bohr$^3$, respectively.}
\label{F-2}
\end{center}
\end{figure*}

The GS electronic structure of oxirane is shown in Fig. \ref{F-2}. The energy spectrum is computed at two levels of theory, LDA (black circles) and \ce{\textit{G}0\textit{W}0} (red squares), and is compared [occupied subspace only in Fig. \ref{F-2}(a)] with experiments (green diamonds). The experimental single-electron energies are sourced from Refs \citenum{Watanabe1957, Basch1969, Schweig1973, Johnson1982, Winkler1999}. \\

The squared KS wavefunction (\textit{charge density}) for each level is shown in Fig. \ref{F-2}. Each of these plots indicates the corresponding KS level which can be tracked by the black dotted lines in the figure. Within the one-shot \textit{GW}A (\ce{\textit{G}0\textit{W}0}), the KS eigenvalues are corrected in a non--self-consistent manner while the KS orbitals approximate the QP wavefunctions \cite{Hybertsen1986}. Therefore, the LDA wavefunction squared plots hereon will be referred to as `quasiparticle levels'. In addition, the labeling for each of the QP levels follows the nomenclature from the LDA (blue-dotted lines in Fig. \ref{F-2}). The two frontier occupied orbitals, highest occupied molecular orbital (HOMO) and HOMO$-1$, exhibit non-bonding $\pi$ (around the O atom) and $\sigma$ characters, respectively, in agreement with other experimental \cite{Watanabe1957, Basch1969, Schweig1973, Johnson1982, Winkler1999} and computational reports \cite{Hirose2006, Grabowsky2008, NIST, Tapavicza2008, Cordova2007, Schmidt2020}. \\

Figures \ref{F-2}(a) and (b) represent the occupied and virtual subspaces, respectively. From Fig. \ref{F-2}(a), it is clearly evident that the QP energies are in close agreement with the experimental values. In Fig. \ref{F-2}(b), the unoccupied KS levels lie below the vacuum level (0 eV). This problem is rectified by \(G_0W_0\) which correctly shifts the corresponding QP energies above the vacuum level.

\subsection{Ionization potential}

From \ce{\textit{G}0\textit{W}0} we obtained the ionization potential (IP) as $10.28 \pm 0.2$ eV [10.48 eV shown in Fig. \ref{F-2}(a)]. The error in estimating the IP value is based on the variation in the results depending on the choice of the reference state; for example, we can evaluate IP using either the GS or the cationic state (removal of an electron). The IP values computed from TOMBO (both LDA and \ce{\textit{G}0\textit{W}0} levels of theory) are compared with tose obtained from experiments and other levels of theory, as shown in Table \ref{tab:IP_comparison}. \\

\begin{table*}[!htbp]
    \centering
    \begin{tabular}{|c|c|c|c|c|c|c|c|c|c|}
        \hline
        Energy & \multicolumn{2}{c|}{TOMBO} & \multicolumn{6}{c|}{Reference \cite{NIST_IP}} & \multirow{2}{*}{Experiment} \cite{Liu1949, Watanabe1957, Lowrey1958} \\
        \cline{2-9}
        (eV) & LDA & \textit{GW}A & PBE & B3LYP & HF & MP4 & CISD & CCSD(T) & \\
        \hline
        IP & 6.55 & $10.28 \pm 0.2$ & 5.79 & 7.27 & 12.12 & 12.11 & 12.15 & 12.10 & 10.565-10.81 \\
        \hline
    \end{tabular}
    \caption{Ionization potential (IP) in eV, computed by LDA and one-shot \textit{GW}A using TOMBO, compared with reference negative-HOMO energy estimates from the listed methods using the 6-31G* basis set \cite{NIST_IP} and experimental ionization energies \cite{Liu1949, Watanabe1957, Lowrey1958}.}
    \label{tab:IP_comparison}
\end{table*}

The computed values from Ref. \citenum{NIST_IP} presented in Table \ref{tab:IP_comparison} are readouts of the negative of the HOMO energy in oxirane obtained from different levels of theory (not exhaustively listed here) such as PBE, hybrid functionals (B3LYP), and wavefunction theories [Hartree-Fock (HF), M{\o}ller-Plesset 4\textsuperscript{th} order perturbation theory (MP4), configuration interaction singles-doubles (CISD), and coupled-cluster singles and doubles with triple excitations perturbatively treated (CCSD(T))] all using the 6-31G* basis set. The negative HOMO energies from the reference calculations are used here as orbital-energy estimates of the IP and compared with the present \(G_0W_0\) result and the experimental values. A wide range of IP estimates is obtained; however, they all differ significantly from the range of 10.57-10.81 eV reported in experiments \cite{Liu1949, Watanabe1957, Lowrey1958}. We note that the reference computed values are basis-set--dependent and that we report only the values obtained with the 6-31G* basis set as it is one of the most widely used basis sets in quantum chemistry calculations. Nevertheless, there is no quantitative agreement between the values obtained from the other basis sets and the experimental reference. Among the orbital-energy estimates listed in Table I, the present \(G_0W_0\) result is closest to the experimental IP, supporting our choice of \(G_0W_0\) for the present electronic-structure calculations.

\subsection{Excited states of oxirane}

The ring-opening mechanism in oxirane is mediated by excited states as has been explored in previous experimental and computational reports \cite{GomerNoyes1950, Takezaki1973, Schmidt2020, Tapavicza2008, Cordova2007, Jenkins2019}. Previous computational reports on oxirane ring-opening have investigated the ring-opening mechanism mediated by the first three singlet excited states: HOMO $\rightarrow$ LUMO (\ce{S1}) (LUMO: lowest unoccupied molecular orbital), HOMO $\rightarrow$ LUMO+1 (\ce{S2}), and HOMO $\rightarrow$ LUMO+2 (\ce{S3}), which involve \textit{n} $\rightarrow$ 3\textit{s}, \textit{n} $\rightarrow$ 3\textit{p\textsubscript{z}}, and \textit{n} $\rightarrow$ $\sigma^*$ transitions, respectively \cite{Tapavicza2008, Cordova2007, Schmidt2020}. The vertical excitation energies (VEEs) reported in these works were computed using a variety of methods: high-quality approaches such as QMC \cite{Tapavicza2008, Cordova2007} and complete active space (CAS) \cite{Cordova2007}, and standard approaches such as  TDDFT/TDA (using different density functional approximations for the time-dependent XC functionals) and occupation-constrained DFT \cite{Schmidt2020}. The VEEs are listed in Table \ref{tab:PAE_comparison}. \\

Within the scope of MBPT, the Bethe--Salpeter equation (BSE) is used to obtain the VEEs based on inputs from the \textit{GW}A \cite{Kuwahara2016, Aoki2019}. However, \textit{GW}A+BSE is heavily sensitive to the initial level of theory used. A conventional \textit{GW}A+BSE procedure involves two steps: solving the QP energies using the QP equations within the \textit{GW}A followed by constructing the BSE kernel using the QP energies. The QP energies are typically solved within the \ce{\textit{G}0\textit{W}0}@LDA framework as they have been demonstrated to be sufficiently accurate \cite{Hybertsen1986}. However, the wavefunctions obtained are within the LDA level. These are used for constructing the excitonic Hamiltonian for the BSE step which can introduce severe evaluation errors \cite{Isobe2018}. The alternative to this approach is to compute VEEs within the scope of \ce{\textit{G}0\textit{W}0}@LDA by using cationic (ground/excited) eigenstates [i.e., $(N-1)$ electron systems] of the many-body Hamiltonian as demonstrated in Refs. \citenum{Isobe2018, Yang2019}. The VEEs are computed by simply subtracting the QP energies of the LUMO of the $(N-1)$ electron system (i.e., $-{\rm IP}$ of the $N$-electron system) from each higher unoccupied level, i.e.,
\begin{eqnarray}
    \Omega_n & = & \epsilon^{(N-1)}_{{\rm LUMO}+n} - \epsilon^{(N-1)}_{{\rm LUMO}} \nonumber \\
    & = & \epsilon^{(N-1)}_{{\rm LUMO}+n} + {\rm IP}^{N}; n \ge 1
\label{eq:PAE_cation}
\end{eqnarray}
The detailed theory behind this approach is provided in Ref. \citenum{Isobe2018} for the interested reader. The main advantage of this approach is that it provides a rigorous one-to-one correspondence between the photoabsorption peaks in the photoabsorption spectrum and the QP energy levels by virtue of the EQPT, which is not available in \ce{\textit{G}0\textit{W}0}@LDA+BSE. Unlike BSE, however, the cationic approach does not provide excitation amplitudes or oscillator strengths. The implicit assumption is that each excited state is characterized by a single electronic configuration. \\

\begin{figure}[!ht]
\begin{center}
\includegraphics[width=\columnwidth]{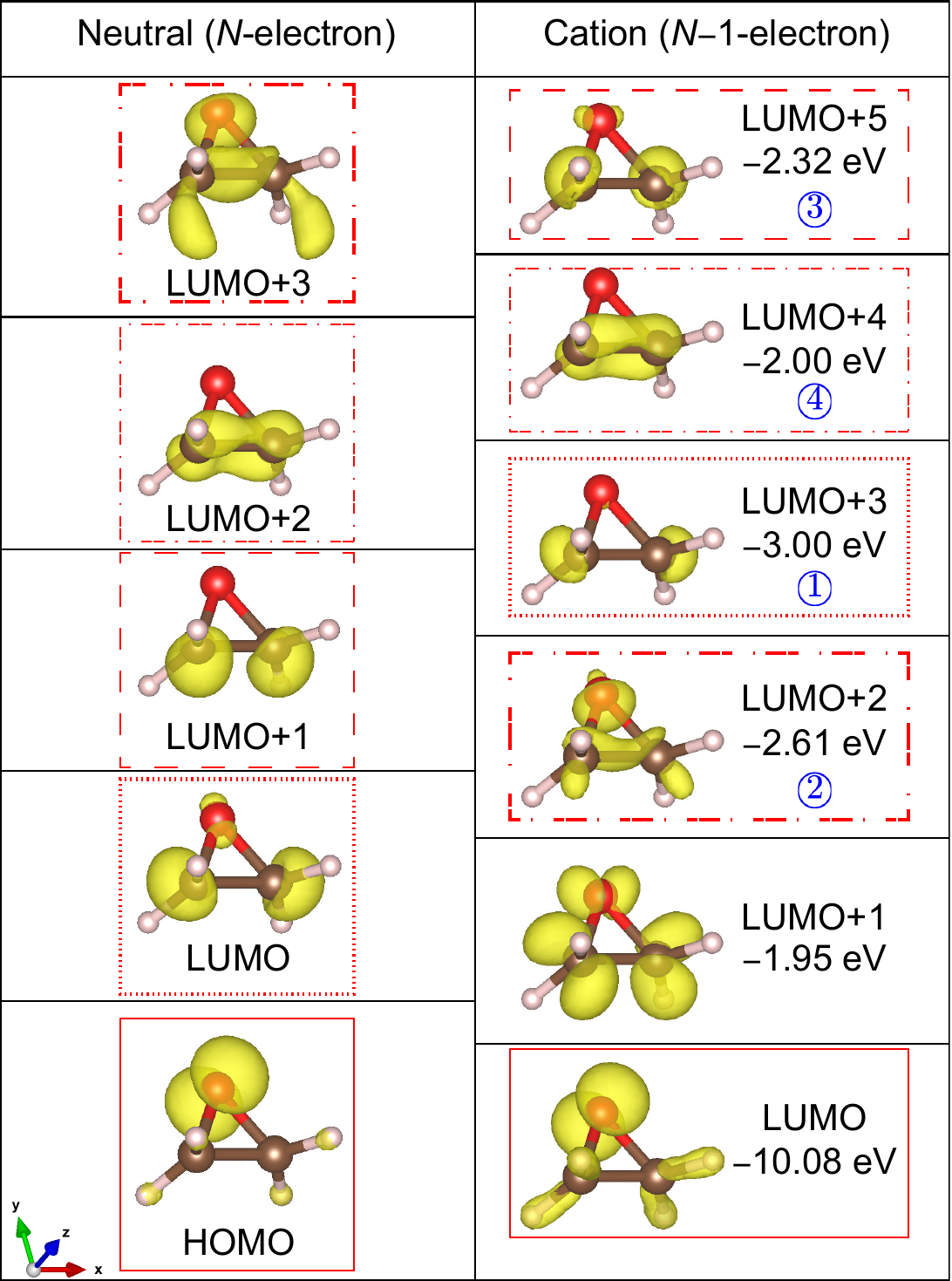}
\caption{A side-by-side comparison of the orbitals obtained from the neutral and cationic references of oxirane. The QP values of the different cationic levels are also provided in the figure. The red boxes with different line styles are guides to the eye to track the cationic orbitals that have the same charge character as the neutral ones, i.e., LUMO (cation) corresponds to HOMO (neutral), LUMO+3 (cation) to LUMO (neutral), etc. The circled numbers in blue represent the order in which the excited states are computed. The viewing orientation is indicated. The isosurfaces for the occupied and unoccupied wavefunction squared plots are $5 \times 10^{-9}$ and $5 \times 10^{-10}$ electrons/bohr$^3$, respectively.}
\label{F-3}
\end{center}
\end{figure}

It should be noted that since \ce{\textit{G}0\textit{W}0} is used, the QP energies obtained do not necessarily need to be ordered in ascending order. For example, this is shown in Fig. \ref{F-3} for the cationic structure of oxirane.
In such a scenario, the lowest $\epsilon^{(N-1)}_{{\rm LUMO}+n}$ value is chosen to compute $\Omega_1$ followed by the next lowest value for $\Omega_2$, etc., while tracking which cationic orbitals match in charge character with those from the neutral orbital set. \\

In Fig. \ref{F-3}, the QP energies of the cationic orbitals are shown along with the order (given by the blue-circled numbers) in which they are considered when computing the VEEs. The red boxes represent the correspondence between the cationic and the neutral orbital sets. Here, LUMO (solid line) from the cationic reference corresponds to HOMO (solid line) from the neutral reference, LUMO+3 (cation, dotted line) to LUMO (neutral, dotted line), LUMO+2 (cation, thick dashed-dotted line) to LUMO+3 (neutral, thick dashed-dotted line), and so on. Based on the ordering of the QP energies and the similarity between corresponding sets of orbitals, we can assign the dominant transitions of the first three excited states of oxirane: HOMO $\rightarrow$ LUMO (\ce{S1}), HOMO $\rightarrow$ LUMO+3 (\ce{S2}), and HOMO $\rightarrow$ LUMO+1 (\ce{S3}). At the outset, this labeling may seem to contradict what has been previously reported in computational studies; however, the charge characters of the orbitals involved in the transitions of the first three excited states are consistent with these reports \cite{Tapavicza2008, Cordova2007, Jenkins2019}. The difference in labeling is primarily due to the usage of the LDA XC functional, while the previous reports used PBE or hybrid functionals such as B3LYP for their ground state calculations. \\

\begin{table}[!htbp]
    \centering
    \begin{tabular}{|c|c|c|c|}
        \hline
        \textbf{Method} & \textbf{{S\textsubscript{1}} (eV)} & \textbf{{S\textsubscript{2}} (eV)} & \textbf{{S\textsubscript{3}} (eV)} \\
        \hline
        Present work & 7.08 & 7.47 & 7.76 \\
        \hline
        TDHF \cite{Cordova2007} & 9.14 & 9.26 & 9.36 \\
        \hline
        TDLDA \cite{Cordova2007} & 6.01 & 6.73 & 6.78 \\
        \hline
        TDB3LYP \cite{Cordova2007} & 6.69 & 7.14 & 7.36 \\
        \hline
        DMC \cite{Tapavicza2008} & $7.51 \pm 0.02$ & $7.62 \pm 0.03$ & $8.64 \pm 0.02$ \\
        \hline
        cDFT \cite{Schmidt2020} & 6.69 & 6.76 & 7.26 \\
        \hline
        Expt. \cite{Liu1949, Lowrey1958, Basch1969} & 7.24 & 7.45 & 7.88 \cite{Liu1949}, 7.89 \cite{Lowrey1958} \\
        \hline
    \end{tabular}
    \caption{VEEs (in eV) of the three low-lying singlet excited states of oxirane computed by \ce{\textit{G}0\textit{W}0}@cation in TOMBO, together with values obtained from other computational approaches and experiments.}
    \label{tab:PAE_comparison}
\end{table}

Table \ref{tab:PAE_comparison} shows that our \ce{\textit{G}0\textit{W}0}@cation results match extremely well with the experimental values (within an energy interval of $\sim0.2$ eV) demonstrating the predictive power of our approach. In spite of the different ordering of the QP energies from our calculation, the nature of the transitions involved in the first three singlet excited states -- Rydberg for \ce{S1} (HOMO $\rightarrow$ LUMO) and \ce{S3} (HOMO $\rightarrow$ LUMO+1), and non-Rydberg \textit{N} $\rightarrow$ \textit{V} for \ce{S2} (HOMO $\rightarrow$ LUMO+3) -- aligns closely with previous reports \cite{Basch1969}.


\subsection{Pre-hop ring-opening dynamics mediated by a HOMO $\rightarrow$ LUMO excitation}



Here, we focus on the ring-opening dynamics initiated by the lowest photoabsorption channel identified in the present calculation, namely, the HOMO $\rightarrow$ LUMO excitation. Dynamics initiated by the higher-energy photoabsorption channels shown in Fig.~\ref{F-3} will be considered in future work. The HOMO\textsubscript{$\alpha$} is predominantly an O-centered nonbonding (\textit{n}) orbital [\textit{b}\textsubscript{1}(\textit{n})] \cite{Tapavicza2008,Li2015}, whereas the LUMO\textsubscript{$\alpha$} exhibits Rydberg \textit{a}\textsubscript{1}(3\textit{s})-like character, as shown in Fig.~\ref{F-3}. The initial excited electronic configuration is constructed by promoting one $\alpha$-spin ($\uparrow$-spin) electron from HOMO\textsubscript{$\alpha$} to LUMO\textsubscript{$\alpha$}, while leaving the $\beta$-spin occupations unchanged. The resulting HOMO $\rightarrow$ LUMO excited configuration is associated with the first singlet excited state \ce{S1},  because the excitation is spin-conserving and originates from the closed-shell singlet ground state. Starting from this electronic configuration at $t = 0$, the coupled electron--nuclear dynamics is propagated using TD\textit{GW}-MD. \\

\begin{figure*}[!ht]
\begin{center}
\includegraphics[width=0.8\textwidth]{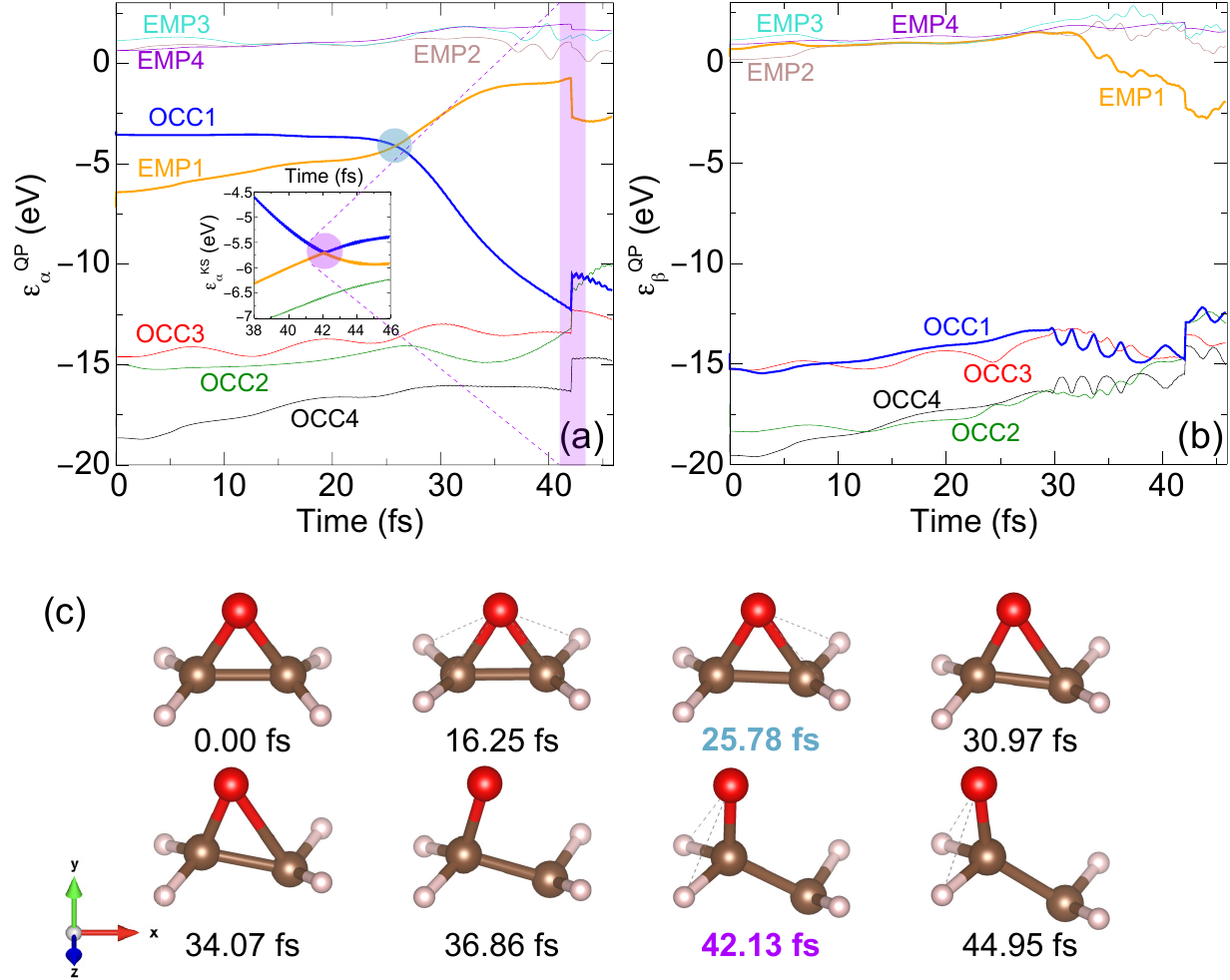}
\caption{The temporal evolution of the QP energies obtained from TD\textit{GW}-MD for the ring-opening dynamics of oxirane mediated by a HOMO $\rightarrow$ LUMO excitation is shown in (a) and (b) for $\alpha$ and $\beta$ spins, respectively. The light-blue shaded circle in (a) corresponds to the crossing between the OCC1\textsubscript{$\alpha$} (thick blue line) and EMP1\textsubscript{$\alpha$} (thick orange line) QP levels. The pink-shaded region in (a) marks the step-like change in the QP energies associated with the avoided crossing (pink circle) between the corresponding OCC1\textsubscript{$\alpha$} (thick blue line) and EMP1\textsubscript{$\alpha$} (thick orange line) KS levels shown in the inset. The trajectory showing the opening of the \ce{C-O-C} ring in oxirane is shown in (c). The times relevant to (QP/KS) level crossings shown in (a) are marked in corresponding colors and are in bold text. The viewing direction is indicated.}
\label{F-4}
\end{center}
\end{figure*}

The temporal evolution of the QP energies for the $\alpha$- and $\beta$-spin ($\downarrow$-spin) channels is shown in Figs.~\ref{F-4}(a) and \ref{F-4}(b), respectively. Representative structural snapshots from the TD\textit{GW}-MD trajectory are shown in Fig.~\ref{F-4}(c).
Following the $\alpha$-spin excitation, the occupations of HOMO\textsubscript{$\alpha$} and LUMO\textsubscript{$\alpha$} are interchanged relative to their ground-state occupations. The now-occupied orbital derived from the ground-state LUMO is denoted OCC1\textsubscript{$\alpha$} [thick blue line in Fig.~\ref{F-4}(a)], whereas the now-empty orbital derived from the ground-state HOMO is denoted EMP1\textsubscript{$\alpha$} [thick orange line in Fig.~\ref{F-4}(a)]. The remaining lower occupied and higher empty levels are labeled OCC$n$ and EMP$n$ ($n > 1$), respectively, according to the energy ordering of their corresponding KS levels. For consistency, the same nomenclature (all occupied orbitals by OCC$n$ and all unoccupied ones by EMP$n$, $n \ge 1$) is used for the $\beta$-spin channel, although its occupation is unchanged by the initial excitation. \\

\begin{figure*}[!ht]
\begin{center}
\includegraphics[width=0.8\textwidth]{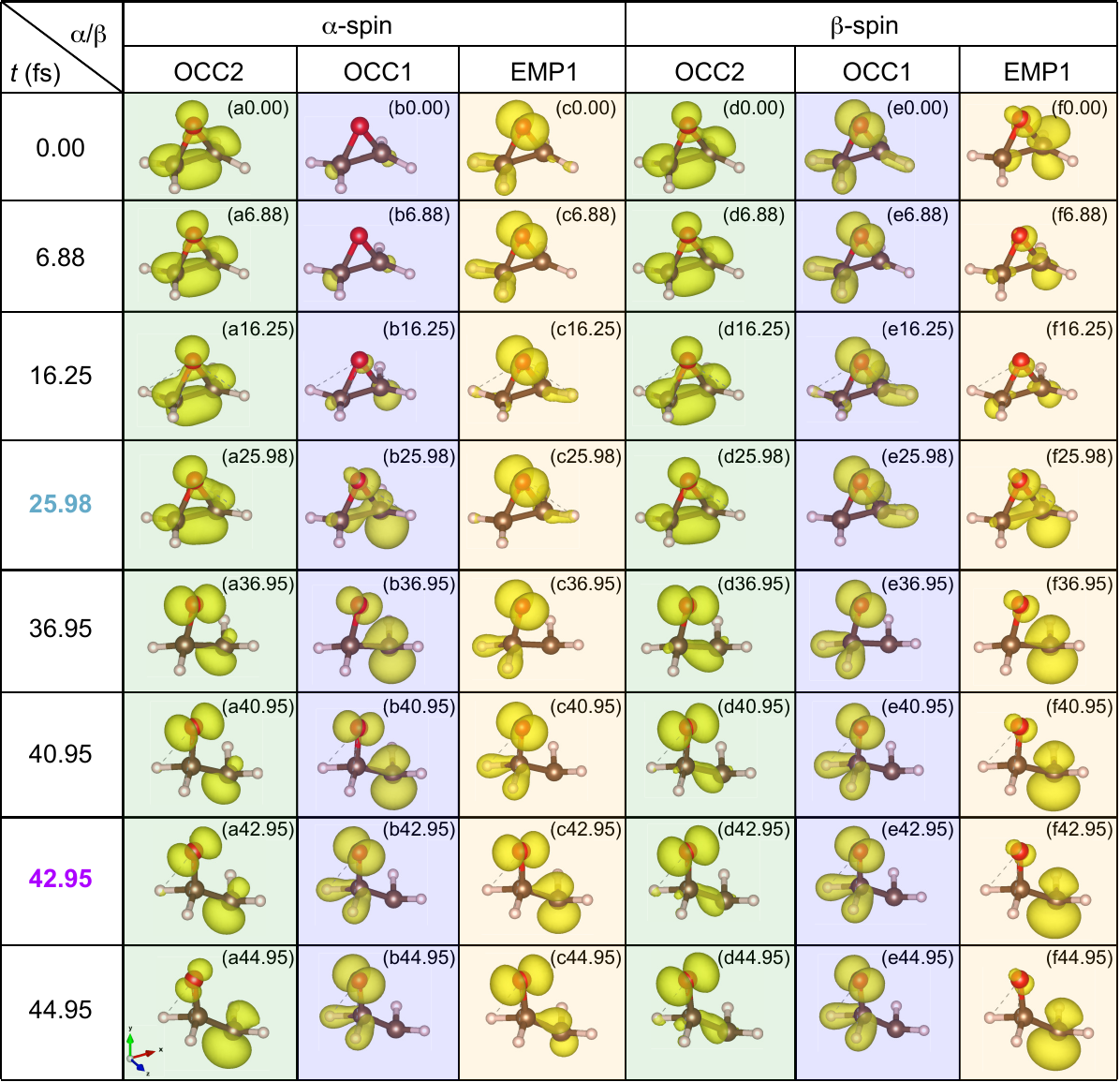}
\caption{The time-evolution of the charge densities of the QP levels of interest for both $\alpha$ and $\beta$ spins is shown for time instants $t = 0, 6.88, 16.25, 25.98, 36.95, 40.95, 42.95$, and $44.95$ fs. The times relevant to (QP/KS) level crossings shown in Fig. \ref{F-4}(a) are shown in the corresponding colors and in bold. The colors of the charge density panels correspond to the colors of the levels in Figs. \ref{F-4}(a) and (b). The isosurface of the charge densities is $5 \times 10^{-9}$ electrons/bohr\textsuperscript{3}. The viewing direction is indicated.}
\label{F-5}
\end{center}
\end{figure*}

The temporal evolution of $\alpha$- and $\beta$-spin charge densities associated with selected frontier QP levels at $t = 0$, 6.88, 16.25, 25.98, 36.95, 40.95, 42.95, and 44.95~fs is shown in Fig. \ref{F-5}.
For the $\alpha$-spin channel, panels (a)--(c) in the charge-density table show OCC2\textsubscript{$\alpha$}, OCC1\textsubscript{$\alpha$}, and EMP1\textsubscript{$\alpha$}, respectively. At the ground-state geometry at $t = 0$, these levels correspond to HOMO$-1$, LUMO, and HOMO, respectively. For the $\beta$-spin channel, panels (d)--(f) show OCC2\textsubscript{$\beta$}, OCC1\textsubscript{$\beta$}, and EMP1\textsubscript{$\beta$}, which correspond to the ground-state HOMO$-1$, HOMO, and LUMO, respectively. Each panel is identified by its letter followed by the simulation time in femtoseconds (fs). \\

Following excitation, the peripheral H atoms undergo out-of-plane motion and progressively approach a geometry in which they are approximately coplanar with the \ce{C-C} bond. Concurrently, the QP energies of OCC1\textsubscript{$\alpha$} and EMP1\textsubscript{$\alpha$} approach each other [Fig.~\ref{F-4}(a)], predominantly because of the rapid increase in the energy of EMP1\textsubscript{$\alpha$}. Surprisingly, EMP1\textsubscript{$\alpha$} largely retains its initial spatial character during the early part of this interval. Changes in the density distribution become apparent as the trajectory approaches $t = 40.95$~fs. At approximately $t = 16.25$~fs, the structure becomes more symmetric as the H atoms pass through a nearly coplanar arrangement [Fig.~\ref{F-4}(c)]. Consequently, the EMP1\textsubscript{$\alpha$} density is approximately related by reflection to that observed at earlier times. Over the same interval, OCC1\textsubscript{$\alpha$} develops increased bonding character [see Figs. \ref{F-5}(b0.00)-(b40.95)]. At $t = 25.78$~fs, the two QP levels OCC1\textsubscript{$\alpha$} and EMP1\textsubscript{$\alpha$} cross each other [light-blue--shaded region in Fig.~\ref{F-4}(a)]. At approximately the same time, the H atoms adopt a \textit{trans}-like arrangement with respect to the \ce{C-C} bond [Fig.~\ref{F-4}(c)]. This structural rearrangement precedes the onset of pronounced asymmetric \ce{C-O} bond elongation. In other words, it is the precursor to the opening of the \ce{C-O-C} ring. It is important to note that the crossing of the two one-particle QP energies does not by itself indicate a completed transition from \ce{S1} to \ce{S0}. This crossing merely reverses the energetic ordering of the two frontier QP levels without changing their occupations. Following the QP-level crossing, the charge characters of OCC1\textsubscript{$\alpha$} and EMP1\textsubscript{$\alpha$} remain unchanged, as illustrated at $t = 36.95$~fs [Figs \ref{F-5}(b36.95) and (c36.95)]. Moreover, the breaking acceleration defined in Section \ref{Method} remains well below the cloning threshold, $\delta_{\rm clone}=3\times10^{-6}$~a.u. Therefore, no surface hop is initiated, and the excited-state TD\textit{GW}-MD propagation is continued. \\

One of the two \ce{C-O} bonds subsequently begins to elongate preferentially, resulting in asymmetric cleavage and opening of the \ce{C-O-C} ring at approximately $t = 36.86$~fs [Fig.~\ref{F-4}(c)]. The asymmetric cleavage and its timescale are consistent with previous oxirane dynamics simulations, which reported completion of the initial ring-opening step within 30-40 fs \cite{Tapavicza2008, Schmidt2020}. Krenz \textit{et al.} also associated ring-opening with the \ce{S1} state, although their trajectories were initialized in \ce{S2} and ring-opening began after an \ce{S2} $\rightarrow$ \ce{S1} hop \cite{Schmidt2020}. This differs from the earlier TDDFT surface-hopping study of Tapavicza \textit{et al.}, in which the \ce{S2} state near the equilibrium geometry was identified as the reactive channel, whereas the lower \ce{S1} state was found to be comparatively unreactive \cite{Tapavicza2008}. \\

At $t = 40.95$~fs, the densities associated with OCC1\textsubscript{$\alpha$} and EMP1\textsubscript{$\alpha$} display mixed O-centered nonbonding and bonding or antibonding character [Figs.~\ref{F-5}(b40.95) and \ref{F-5}(c40.95)]. This mixed character indicates that OCC1\textsubscript{$\alpha$} and EMP1\textsubscript{$\alpha$} no longer retain their original orbital characters as the \ce{C-O} bond elongates. At approximately $t = 42.13$~fs, the QP energies exhibit a step-like change [pink-shaded region in Fig.~\ref{F-4}(a)]. This arises from an avoided crossing between the two frontier KS levels OCC1\textsubscript{$\alpha$} and EMP1\textsubscript{$\alpha$}, as shown in the inset of Fig.~\ref{F-4}(a). The KS levels follow the same labeling and color convention as the corresponding QP levels. The accompanying exchange of orbital character is visible in the OCC1\textsubscript{$\alpha$} and EMP1\textsubscript{$\alpha$} densities at $t = 42.95$~fs, just after this avoided crossing. The avoided crossing and exchange of orbital character indicate that the trajectory has reached a region in which de-excitation to a lower electronic state becomes possible. Within the present algorithm, the decision to initiate a surface hop is made using the breaking-acceleration criterion rather than the level crossing alone. At this stage, the breaking acceleration exceeds $\delta_{\rm clone}$, identifying this region as the first candidate for a surface hop. \\

Ideally, the post-hop trajectory would be initiated at this time. However, a stable GS SCF solution could not be obtained at the corresponding geometry because of the near-degeneracy of OCC1\textsubscript{$\alpha$} and EMP1\textsubscript{$\alpha$}. This difficulty is consistent with previous electronic-structure descriptions of ring-opened oxirane, in which biradicaloid (two frontier electrons are associated predominantly with the O and the C2 atoms) and zwitterionic (an electron is transferred toward O from C2, producing a contribution represented schematically as \ce{O^{-}-CH2-H2C+}) electronic configurations become closely competing upon \ce{C-O} bond cleavage \cite{GomerNoyes1950, Tapavicza2008, HuixRotllant2010}. The resulting open-shell character makes the GS electronic structure difficult to describe using a conventional single-reference closed-shell solution \cite{Cordova2007, Tapavicza2008, HuixRotllant2010}. The surface hop was therefore deferred to approximately $t = 44.95$~fs, for which the molecular structure and charge densities are shown in Figs.~\ref{F-4}(c) and \ref{F-5}(a44.95)-(f44.95), respectively. \\


\subsection{Surface hopping and post-hop dynamics}

The ring-opening motion observed here corresponds to the initial step of the Gomer-Noyes mechanism proposed from early oxirane photolysis experiments and subsequently examined in other experimental and computational studies \cite{GomerNoyes1950, Takezaki1973, Tapavicza2008, Li2015, Schmidt2020}. In this mechanism, ring opening is followed by intramolecular H transfer and subsequent fragmentation. Previous surface-hopping studies found that the initial ring opening occurs on an excited electronic state, whereas H transfer typically takes place after population transfer to the electronic ground state \cite{Tapavicza2008, Schmidt2020}. However, further fragmentation was observed to occur less frequently \cite{Schmidt2020} than in experiments \cite{GomerNoyes1950, Takezaki1973}. \\

Here, we examine the post-hop dynamics initiated from the ring-opened geometry at $t = 44.95$~fs using two electronic configurations: an open-shell singlet GS and the triplet GS, i.e., the lowest-energy electronic solution within the triplet spin manifold. The triplet configuration is considered because its open-shell occupation can be represented straightforwardly by a single KS determinant with parallel-spin electrons occupying the two frontier orbitals. This trajectory provides a computationally well-defined open-shell pathway but should not be interpreted as evidence that intersystem crossing to the triplet manifold is the actual relaxation mechanism of oxirane. We first discuss the triplet-GS dynamics, followed by the open-shell singlet-GS dynamics. \\

Following the surface hop, the nuclear motion is propagated using spin-polarized GS Born-Oppenheimer molecular dynamics (BOMD), with the electronic occupations initialized for the selected post-hop configuration. This treatment is consistent with the TSH framework, in which the nuclei evolve according to the BO forces of the active electronic state between hopping events \cite{Tully1990, Tavernelli2010, Tapavicza2008, Curchod2018}. No subsequent hops between electronic states are included in the present trajectories. \\

At $t = 44.95$~fs, the nuclear velocities obtained from the pre-hop TD\textit{GW}-MD trajectory were rescaled along the nuclear linear-momentum direction to conserve the total energy, as described in the METHODS section. For each downward transition, the decrease in electronic energy was transferred to the nuclear kinetic energy. The ring-opened geometry and the rescaled velocities were then used as the initial conditions for the triplet and open-shell singlet BOMD trajectories.

\subsubsection{Post-hop dynamics on the triplet ground state}

For the triplet post-hop trajectory, the two frontier electrons were assigned parallel spins, giving 13 occupied $\alpha$-spin orbitals and 11 occupied $\beta$-spin orbitals. All the energy levels of the $\alpha$ and $\beta$ spin orbitals are now relabeled, for example, pre-hop OCC1\textsubscript{$\alpha$} as $^3$OCC2\textsubscript{$\alpha$}, EMP1\textsubscript{$\alpha$} as $^3$OCC1\textsubscript{$\alpha$}, OCC1\textsubscript{$\beta$} as $^3$EMP1\textsubscript{$\beta$}, OCC2\textsubscript{$\beta$} as $^3$OCC1\textsubscript{$\beta$}, etc. The ``3'' in the superscript denotes that the levels now correspond to the triplet electronic state. The ring-opened geometry at $t = 44.95$~fs and the velocity-rescaled nuclear momenta described above were used to initiate spin-polarized ground-state BOMD. The trajectory was propagated with a nuclear time step of $0.1$~fs. \\

\begin{figure*}[!ht]
\begin{center}
\includegraphics[width=0.8\textwidth]{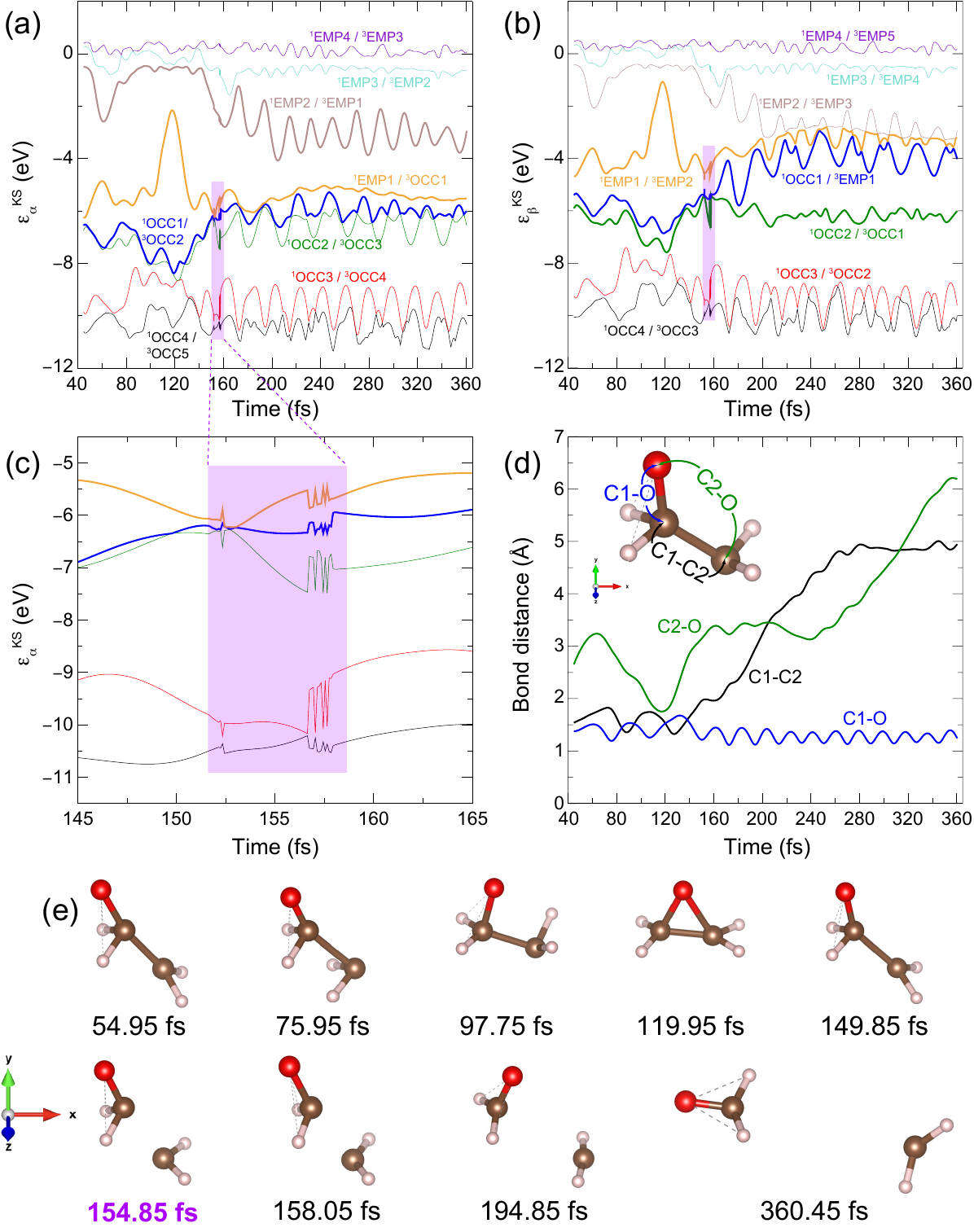}
\caption{Post-hop dynamics on the triplet ground state initiated from the ring-opened geometry at $t = 44.95$~fs. The temporal evolution of selected KS eigenvalues obtained during spin-polarized ground-state BOMD is shown for the (a) $\alpha$- and (b) $\beta$-spin channels. The notation $^{1}{\rm X}/^{3}{\rm Y}$ gives the pre-hop singlet and post-hop triplet labels of the same tracked level. Panel (c) shows an expanded view of selected $\alpha$-spin KS eigenvalues between $t = 145$ and 165~fs. The pink-shaded regions in (a)-(c) mark the interval $t = 151.85$-158.05~fs, during which several frontier levels undergo abrupt changes associated with a reordering and redistribution of orbital character. The evolution of the \ce{C-C} and \ce{C-O} bond distances is shown in (d). Representative structures along the trajectory are shown in (e); the structure at $t = 154.85$~fs, corresponding to the electronic-reorganization interval, is highlighted in purple. The viewing direction is indicated.}
\label{F-6}
\end{center}
\end{figure*}

\begin{figure*}[!ht]
\begin{center}
\includegraphics[width=0.8\textwidth]{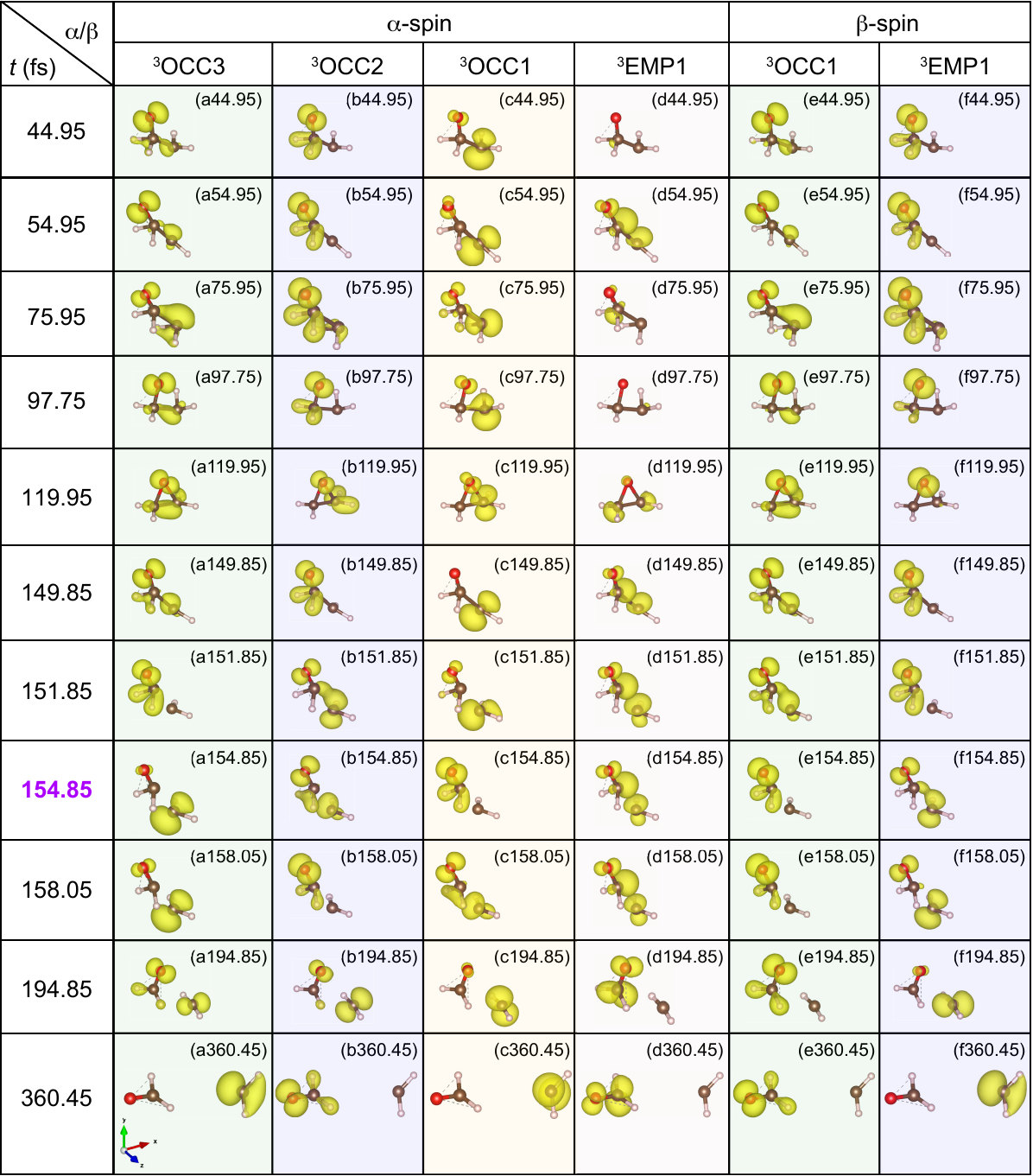}
\caption{Temporal evolution of the charge densities of selected KS levels for both $\alpha$ and $\beta$ spins during the post-hop triplet-ground-state BOMD trajectory. The rows correspond to the sampled times $t = 44.95$, 54.95, 75.95, 97.75, 119.95, 149.85, 151.85, 154.85, 158.05, 194.85, and 360.45~fs. The time relevant to the significant electronic reorganization ($t = 154.85$ fs) in Figs. \ref{F-6}(a)-(c) is highlighted in purple. The colors of the charge density panels correspond to the colors of the levels in Figs. \ref{F-6}(a) and (b). The isosurface value of the charge densities is $5 \times 10^{-9}$ electrons/bohr\textsuperscript{3} (with the exception of the isosurface value for the charge density of \textsuperscript{3}EMP1\textsubscript{$\alpha$/$\beta$} for $t = 119.95$ fs which was set to $5 \times 10^{-10}$ electrons/bohr\textsuperscript{3}). The viewing direction is indicated.}
\label{F-7}
\end{center}
\end{figure*}

Figure~\ref{F-6} summarizes the post-hop triplet-GS BOMD trajectory initiated from the ring-opened geometry at $t = 44.95$~fs. 
Because the present BOMD implementation in TOMBO does not invoke a one-shot \textit{GW} calculation at every MD step, the temporal evolution of the electronic structure is represented by the instantaneous spin-resolved KS eigenvalues [Figs.~\ref{F-6}(a) and \ref{F-6}(b)]. The corresponding evolution of selected frontier-level charge densities is shown in Fig.~\ref{F-7}. \\

Immediately after the hop, the ring opening continues with the \ce{C{2}-O} bond distance increasing [green line in Fig. \ref{F-6}(d)]. There is also a simultaneous increase in the \ce{C{1}-C{2}} bond length [black line in Fig. \ref{F-6}(d)]. During this stage, the selected KS frontier-orbital densities remain mostly localized on different sub-fragments of the ring-opened geometry [Figs.~\ref{F-7}(a44.95)-(f44.95)]. This downward motion of the C2 atom away from O continues for nearly 30 fs, forming a nearly linear O-C1-C2 arrangement [see for example, $t = 54.95$ fs in Fig. \ref{F-6}(e)] before the \ce{C{2}-H{3}} and \ce{C{2}-H{4}} bonds bend inward to produce an `armchair'-like structure at $t = 75.95$ fs [Fig. \ref{F-6}(e)]. At this stage, the charge densities become more delocalized and are distributed over the distorted geometry [Figs. \ref{F-7}(a54.95)-(f54.95) and (a75.95)-(f75.95)]. \\

As time progresses, the O atom approaches C1, shortening the \ce{C{1}-O} bond. As C1 recoils from O, it intermittently pushes the C2 atom away from C1. The O atom then bounces off the C1 atom and begins to move away from it, while pushing C1 closer to C2. However, the strong recoil of the C1 atom from O pushes the C2 atom away, which induces a reversal of the motion of the H3 and H4 atoms bonded to it at $t = 97.75$ fs. The nature of the charge distribution of all the levels at this time [Figs. \ref{F-7}(a97.75)-(f97.75)] closely resembles that at $t = 44.95$ fs [Figs. \ref{F-7}(a44.95)-(f44.95)]. \\

The C2-H3-H4 fragment continues its upward motion due to the residual recoil from the previous process, while the O atom continues to move away from C1. At $t = 119.95$ fs, the \ce{C{1}-O} bond length reaches its maximum while the \ce{C{2}-O} bond distance reaches its minimum due to the persistent upward motion of C2-H3-H4. Here, the previously dissociated C2 atom is brought back into bonding proximity with O and transiently restores an oxirane-like three-membered-ring geometry [Fig. \ref{F-6}(e)]. Unusually large peaks are observed for the orange-colored KS levels $^1$EMP1\textsubscript{$\alpha$} / $^3$OCC1\textsubscript{$\alpha$} and $^1$EMP1\textsubscript{$\beta$} / $^3$EMP2\textsubscript{$\beta$} in Figs. \ref{F-6}(a) and (b), respectively, which coincide with this transient oxirane-like formation. There is a weak resemblance between the charge distribution of the frontier levels at $t = 119.95$ fs [Figs. \ref{F-7}(a119.95)-(f119.95)] and that for the GS oxirane structure. However, it is important to point out that the KS energies for this structure do not match those of the initial GS structure. Consequently, this transient ring-like geometry at $t = 119.95$ fs should not be interpreted as recovery of the initial closed-shell singlet state of oxirane. This re-closure is short-lived, i.e., the residual nuclear kinetic energy carried by the post-hop trajectory drives the system away from the transiently ring-closed configuration, and the \ce{C{2}-O} bond elongates again, causing the ring to reopen. \\

The geometry at $t = 149.85$ fs, upon reopening, is similar to that early in the simulation ($t = 54.95$ fs). This can also be seen from the nearly one-to-one correspondence between the two sets of KS energies at these two time instants. Visually, the charge characters of all levels at $t = 149.85$ fs [Figs. \ref{F-7}(a149.85)-(f149.85)] resemble closely those at $t = 54.95$ fs. Immediately after this time, there is a reorganization in the electronic structure as shown by abrupt changes in the KS eigenvalues [pink shaded regions in Figs. \ref{F-6}(a) and (b), and in the expanded view for $\alpha$ KS energies in (c)]. This occurs in the $t \in [150, 160]$ fs interval. During this time interval, the $\beta$-spin frontier occupied and empty KS orbitals, $^3$OCC1\textsubscript{$\beta$} [thick green line in Fig. \ref{F-6}(b)] and $^3$EMP1\textsubscript{$\beta$} [thick blue line in Fig. \ref{F-6}(b)], respectively, approach each other and exhibit near degeneracy at around $t = 154.85$ fs. At $t = 151.85$ fs, before the near-degenerate region, $^3$EMP1\textsubscript{$\beta$} has a localized charge population around the formaldehyde-like sub-fragment [Fig. \ref{F-7}(f151.85)] while $^3$OCC1\textsubscript{$\beta$} has a more delocalized charge character [Fig. \ref{F-7}(e151.85)]. As the two levels approach near-degeneracy at $t = 154.85$ fs, they exchange their spatial characters [Figs. \ref{F-7}(e154.85) and (f154.85)]. However, it should be noted that the charge densities obtained here are a result of an unconverged SCF process owing to this energy degeneracy. The character exchange is more clearly resolved in the converged orbitals after this interval [see for example, Figs. \ref{F-7}(e158.05) and (f158.05)]. Orbitals that previously exhibited the charge density spread across the two carbon centers now evolve toward fragment-localized states. We therefore interpret this interval as a concerted spin-resolved electron redistribution between the emerging \ce{H2CO}-like and \ce{CH2}-like fragments. \\

This electronic redistribution coincides with the onset of irreversible \ce{C-C} cleavage. The \ce{C-C} distance, which is approximately $2$~\AA\ in the reorganization region, increases continuously thereafter [Fig.~\ref{F-6}(d)]. It exceeds $3$~\AA\ by approximately $t = 195$~fs and approaches $5$~\AA\ near $t = 280$~fs, after which the two carbon-containing fragments remain separated. The structures at $t = 194.85$ and 360.45~fs illustrate this progressive fragmentation [Fig.~\ref{F-6}(e)]. Simultaneously, the KS charge densities at $t = 194.85$ and 360.45~fs become predominantly localized on one or the other fragment [Figs.~\ref{F-7}(a194.85)-(f194.85) and (a360.45)-(f360.45)], confirming the breaking of the original \ce{C-C} bond. \\

\begin{figure*}[!ht]
\begin{center}
\includegraphics[width=0.9\textwidth]{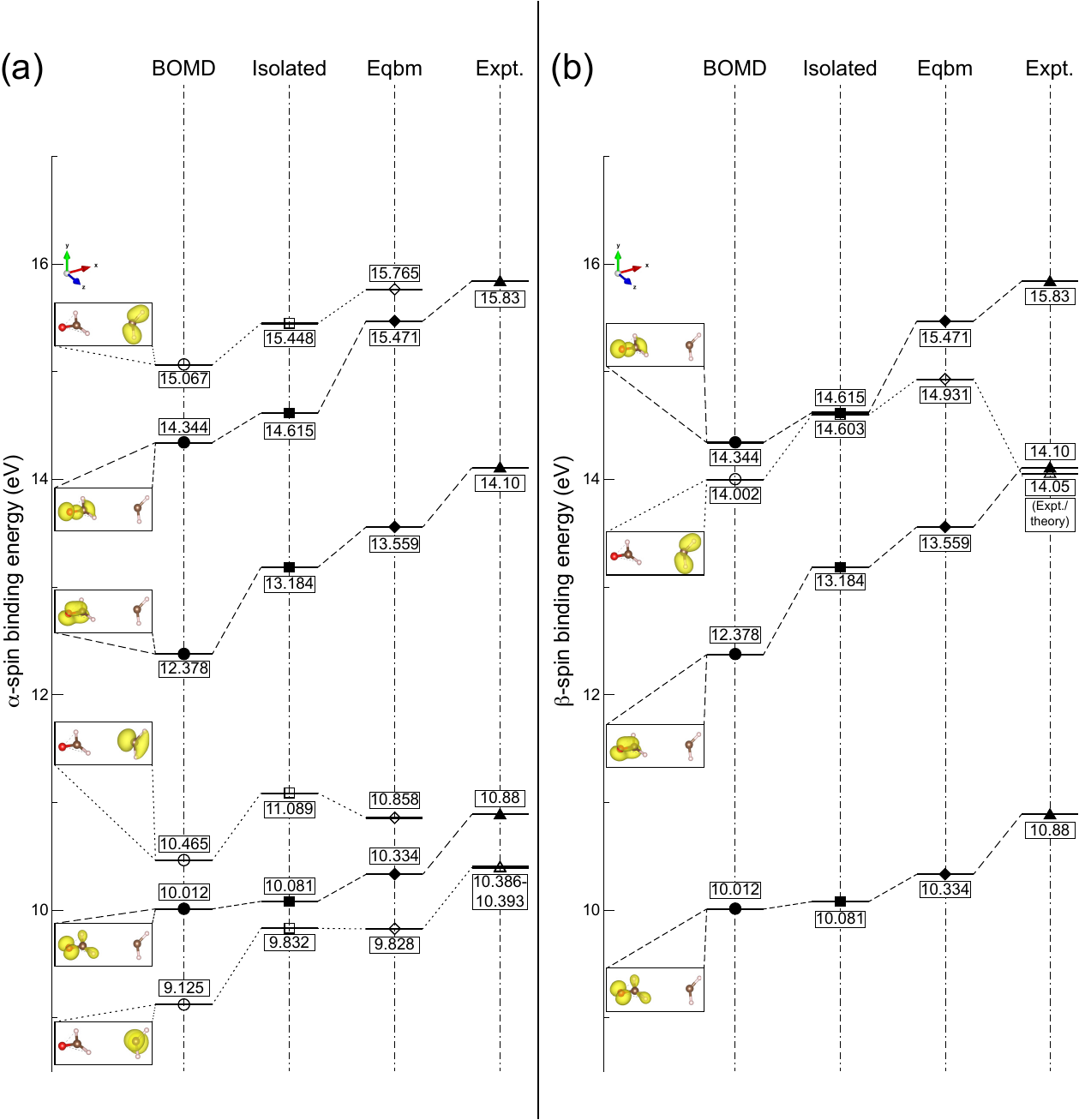}
\caption{Comparison of one-shot \textit{GW} QP binding energies, $-\varepsilon^{\mathrm{QP}}$, associated with the formaldehyde and methylene fragments formed during the post-hop triplet-ground-state BOMD trajectory. Panels (a) and (b) show the $\alpha$- and $\beta$-spin channels, respectively. Filled symbols denote states localized on the formaldehyde (\ce{H2CO}) fragment, whereas open symbols denote states localized on methylene (\ce{CH2}). For each state, QP binding energies obtained from the full BOMD structure at $t = 360.45$~fs (labeled by circles) are compared with those obtained for the corresponding isolated fragment at the geometry extracted from the same BOMD snapshot (labeled by squares) and at its equilibrium geometry (labeled by diamonds). Representative fragment-localized charge densities from the full BOMD structure are shown alongside the corresponding levels, and dotted/dashed lines connect states with matching orbital character. Available gas-phase photoelectron and photoionization energies are shown for comparison \cite{BrundleTurner1967, Baker1968, Niu1993, Litorja1998, Willitsch2002, Coudert2018}. For triplet methylene, the experimental first ionization energy corresponds to formation of the lowest doublet state of \ce{CH2+}; the higher-energy reference near $14.05$~eV [marked as `(Expt./theory)'] is a theoretical estimate for formation of the $\tilde{a}^{+}\,{}^{4}A_{2}$ cationic state based on the calculated doublet--quartet separation \cite{Brinkmann2002}, rather than a direct experimental measurement. The viewing directions are indicated.}
\label{F-8}
\end{center}
\end{figure*}

Based on the fragment-localized charge densities at $t = 360.45$ fs, the resulting products are assigned as formaldehyde, \ce{H2CO}, and methylene, \ce{CH2}. This assignment was further examined by performing \ce{\textit{G}0\textit{W}0} calculations on the full BOMD structure at $t = 360.45$~fs and comparing the resulting QP binding energies ($-\varepsilon^{\mathrm{QP}}$) with those obtained for the isolated fragments at the geometries extracted from the same BOMD snapshot as well as at their respective equilibrium geometries (Fig.~\ref{F-8}).
The $\alpha$- and $\beta$-spin channels are shown separately in Figs.~\ref{F-8}(a) and (b), respectively. Filled symbols denote states localized on the \ce{H2CO} fragment, whereas open symbols denote states localized on \ce{CH2}. Representative charge densities from the full BOMD structure are shown alongside the corresponding levels to establish the fragment assignment, while the connecting lines (dotted for \ce{CH2} and dashed for \ce{H2CO}) follow states with the same orbital character through the isolated-fragment and equilibrium structure calculations. \\


The QP energies of the formaldehyde- and methylene-localized states are on comparable energy scales with their extracted isolated-fragment and equilibrium counterparts, although quantitative shifts of several tenths to about 1 eV are present because the fragments in the BOMD configuration are vibrationally distorted and remain embedded in the electrostatic environment of the complete system. For instance, for the formaldehyde-localized states, the binding energies of the full BOMD structure are 10.012, 12.378, and 14.344 eV in both spin channels, while for the equilibrium geometry, these are 10.334, 13.559, and 15.471 eV. The experimentally observed valence-ionization energies $10.8$-$10.9$, $14.10$, and $15.83$~eV reported in gas-phase photoelectron measurements \cite{BrundleTurner1967, Baker1968, Niu1993} differ significantly from the BOMD values for the precise reasons mentioned above, but are close to the values for the equilibrium geometry. \\


The methylene-localized levels display the expected spin dependence of the open-shell triplet fragment. In the $\alpha$-spin channel, the calculated frontier QP binding energy changes from $9.13$~eV in the full BOMD structure to $9.83$~eV for isolated \ce{CH2} at both the extracted and equilibrium geometries. This level is most naturally compared with the experimentally measured first ionization of triplet methylene to the lowest doublet state of \ce{CH2+}. Independent photoionization and rotationally resolved photoelectron measurements place this threshold at $10.393\pm0.011$~eV and $10.3864\pm0.0004$~eV, respectively \cite{Litorja1998, Willitsch2002, Coudert2018}. \\

On the other hand, for the $\beta$-spin channel, the corresponding QP binding energy evolves from 14.002 eV in the full BOMD structure to 14.603 eV at the extracted isolated-fragment geometry, and 14.931 eV at the equilibrium geometry. Photoelectron experiments do not provide a directly measured ``$\beta$-spin ionization energy'' because the observed transitions are classified according to the final electronic states of \ce{CH2+}, rather than according to the spin label of the removed orbital. The relevant high-spin comparison is therefore the theoretically predicted low-lying $\tilde{a}^{+},{}^{4}A_{2}$ state of \ce{CH2+}, which lies approximately $3.66$~eV above the $\tilde{X}^{+},{}^{2}A_{1}$ cation ground state \cite{Brinkmann2002}. Combining this separation with the experimental first ionization threshold gives an estimated removal energy of approximately $14.05$~eV for access to the quartet cationic manifold. This value is shown only as a theoretical reference in Fig.~\ref{F-8}, labeled by `(Expt./theory)'; no direct experimental term energy for this quartet state appears to be available. \\

The close correspondence of the fragment-localized orbital shapes, their QP binding-energy change upon isolation of the fragments, and the agreement of the equilibrium-fragment values with available experimental ionization energies together provide an independent validation of the assignment of the dissociation products as \ce{H2CO + ^3CH2}. The formation of this product pair is also consistent with previous studies of the triplet \ce{CH2CH2O} biradical, for which direct dissociation to \ce{H2CO + ^3CH2} has been identified as an accessible triplet reaction channel \cite{Hu2008}, as well as with vacuum-ultraviolet photolysis experiments on oxirane reporting \ce{CH2 + HCHO} among the primary dissociation products \cite{Takezaki1973}. Thus, conditional on population transfer to the triplet manifold, the present trajectory follows a direct \ce{C-C}-cleavage pathway to \ce{H2CO + CH2}, rather than the H-transfer pathway characteristic of the Gomer-Noyes mechanism on the singlet ground-state surface \cite{GomerNoyes1950, Tapavicza2008, Schmidt2020}.

\subsubsection{Post-hop dynamics on the open-shell singlet ground state}

We next consider the post-hop dynamics on the singlet ground-state surface. At $t=44.95$~fs, the ring-opened structure was initialized in a broken-symmetry open-shell singlet configuration, with oppositely oriented spin densities predominantly localized on the O and C2 atoms. The rescaled nuclear velocities described above were then used to initiate spin-polarized GS-BOMD. As in the triplet trajectory, the temporal evolution of the electronic structure is represented by the instantaneous spin-resolved KS eigenvalues, since a one-shot \textit{GW} calculation is not performed at every BOMD step. \\

\begin{figure*}[!ht]
\begin{center}
\includegraphics[width=0.8\textwidth]{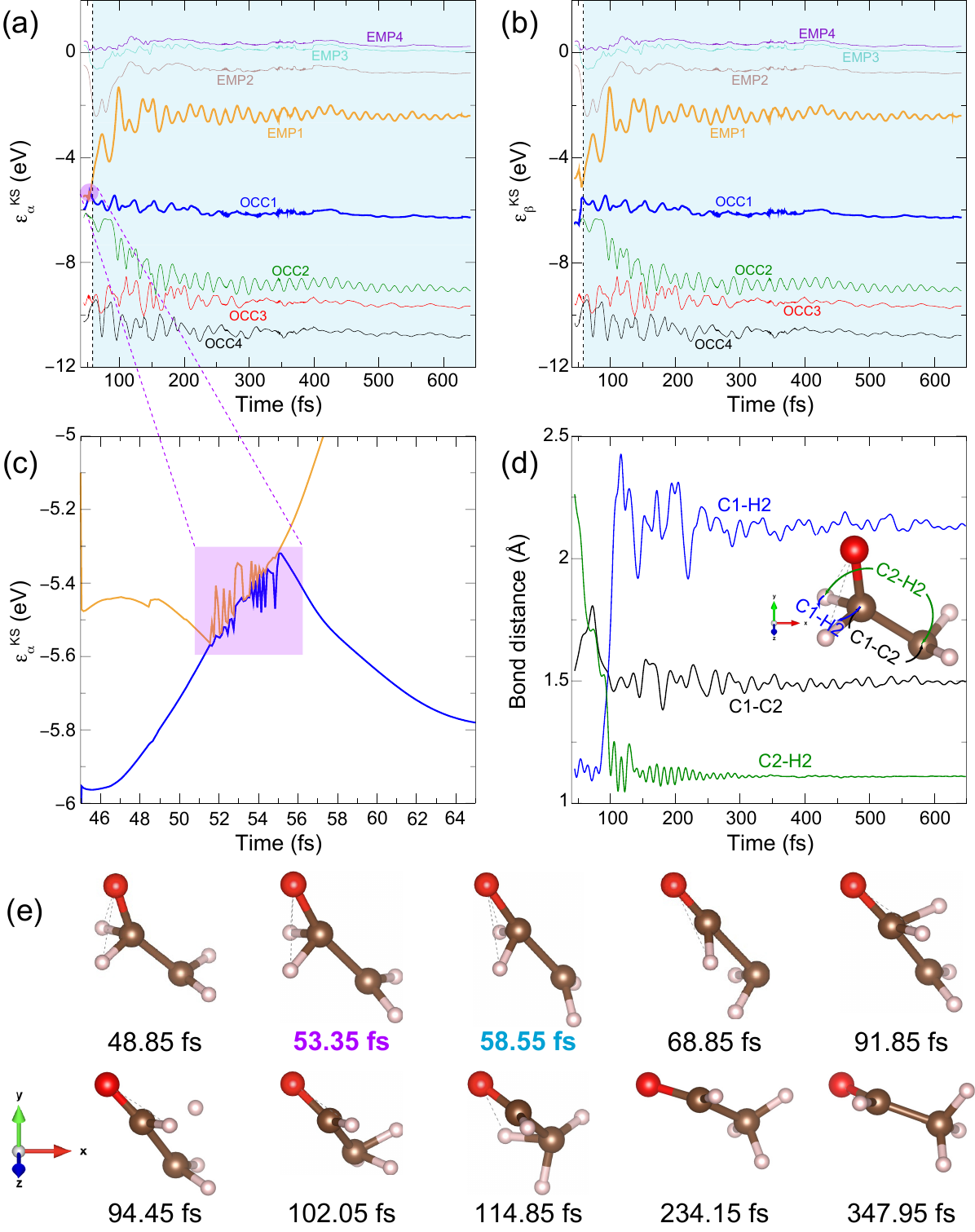}
\caption{Post-hop dynamics on the open-shell singlet ground state initiated from the ring-opened geometry at $t=44.95$~fs. The temporal evolution of selected instantaneous KS eigenvalues obtained during spin-polarized ground-state BOMD is shown for the (a) $\alpha$- and (b) $\beta$-spin channels. The vertical black dashed lines in (a) and (b) mark the onset of the spin-symmetry-preserved region at $t=58.55$~fs. Panel (c) shows an enlarged view of the $\alpha$-spin OCC1 and EMP1 levels immediately following the hop; the shaded region marks the interval over which the frontier electronic structure undergoes rapid reorganization. Panel (d) shows the temporal evolution of the \ce{C{1}-H{2}}, \ce{C{2}-H{2}}, and \ce{C{1}-C{2}} distances. The inset defines the atom labels used for these distances. Representative structures along the trajectory are shown in (e). The geometries at $t=53.35$ and 58.55~fs highlight the near-degenerate frontier region and the onset of the spin-symmetry-preserved electronic structure, respectively. The viewing directions are indicated.}
\label{F-9}
\end{center}
\end{figure*}

\begin{figure*}[!ht]
\begin{center}
\includegraphics[width=0.8\textwidth]{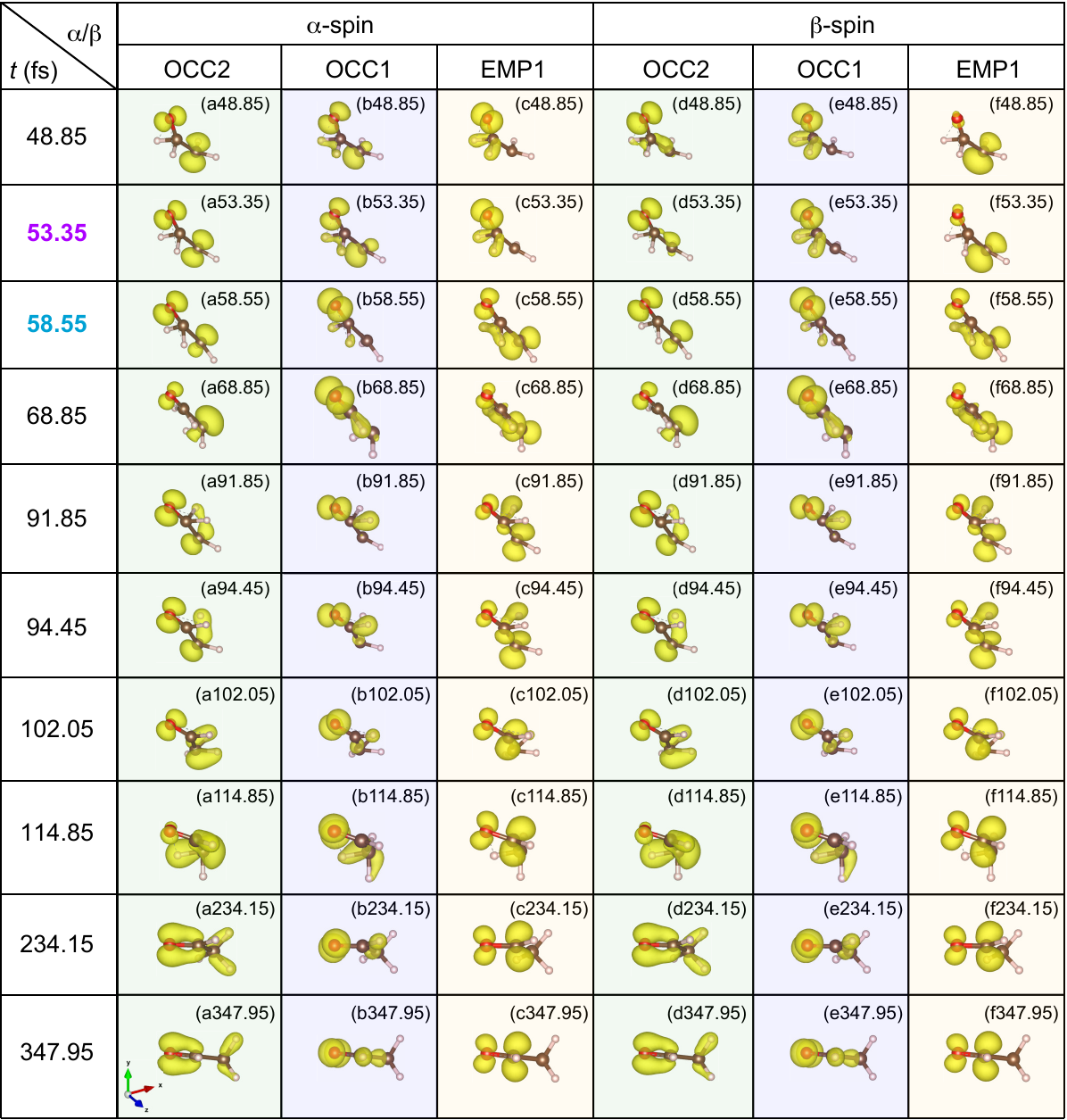}
\caption{Temporal evolution of selected frontier charge densities during the post-hop open-shell singlet ground-state BOMD trajectory. The columns show corresponding $\alpha$- and $\beta$-spin orbital pairs, while the rows correspond to selected geometries along the trajectory at $t=48.85$, 53.35, 58.55, 68.85, 91.85, 94.45, 102.05, 114.85, 234.15, and 347.95~fs. The times associated with the near-degeneracy of the $\alpha$-spin HOMO and LUMO (at $t = 53.35$ fs), and the onset of spin-symmetry (at $t = 58.55$ fs) are marked as in Fig. \ref{F-9}(e). The colors of the charge density panels correspond to the colors of the levels in Figs. \ref{F-9}(a) and (b). The isosurface value of the charge densities is $5 \times 10^{-9}$ electrons/bohr\textsuperscript{3}. The viewing direction is indicated.}
\label{F-10}
\end{center}
\end{figure*}

The temporal evolution of the $\alpha$- and $\beta$-spin KS eigenvalues is shown in Figs.~\ref{F-9}(a) and
\ref{F-9}(b), respectively. The corresponding evolution of selected frontier-level charge densities is shown in Fig.~\ref{F-10}.
Immediately after the surface hop, the two spin channels remain inequivalent, reflecting the broken-symmetry open-shell character imposed at the ring-opened geometry. This broken-spin-symmetry is apparent from the corresponding orbital charge densities at $t=48.85$~fs (Figs.~\ref{F-10}(a48.85)-(f48.85)), where several corresponding $\alpha$- and $\beta$-spin orbitals possess different spatial characters. \\

During the first $\sim10$ fs following the hop, the ring further opens up leading to stronger open-shell character. This is accompanied by a rapid reorganization of the frontier electronic structure. Figure~\ref{F-9}(c) shows an enlarged view of the $\alpha$-spin OCC1 and EMP1 levels over this interval. The two levels approach one another and become essentially degenerate at $t = 53.35$~fs, with KS eigenvalues of $-5.423$ and $-5.422$~eV, respectively. The SCF convergence at this geometry is not achieved because of this near-degeneracy. Nevertheless, this identifies the interval in which the initially stabilized broken-symmetry configuration undergoes a pronounced change in the electronic structure. \\

At $t = 58.55$~fs, the O, C1, C2, H3, and H4 atoms together form an armchair-like configuration. The charge characters of OCC1\textsubscript{$\alpha$} and EMP1\textsubscript{$\alpha$} swap compared to those at $t = 48.85$ and 53.35 fs [Figs. \ref{F-10}(b48.85)-(c48.85) and (b53.35)-(c53.35)], indicating an actual crossing between the two levels as a result of the near-degeneracy. Due to this reorganization of the frontier orbitals in the $\alpha$-spin channel, the corresponding orbitals in the two spin channels acquire nearly identical spatial character [Figs. \ref{F-10}(a58.55)-(f58.55)], in marked contrast to the spin-asymmetric distributions present immediately after the hop, implying that the spin-symmetry is ``restored''. This pairwise correspondence persists at later sampled geometries. We therefore identify $t\approx58.55$~fs as the onset of the spin-symmetry-preserved region of the trajectory. In electronic-structure terms, the initially broken-symmetry biradicaloid KS solution evolves into a spin-paired singlet solution as the molecular geometry moves away from the strongly open-shell ring-opening region. \\

Following the restoration of spin symmetry, the intramolecular H-transfer step is initiated. Immediately after the hop, H2 remains bonded to C1, with $r_{\ce{C{1}-H{2}}}$ close to a normal \ce{C-H} bond length of about 1.11~\AA~[Fig. \ref{F-9}(d)], whereas $r_{\ce{C{2}-H{2}}}$ is substantially larger (around 2.25~\AA, green line). During the subsequent dynamics, these two distances evolve in opposite directions. Near $t\approx90$-100~fs, $r_{\ce{C{1}-H{2}}}$ increases rapidly while $r_{\ce{C{2}-H{2}}}$ contracts, demonstrating transfer of H2 from C1 to C2. The two distances change almost simultaneously, as shown in Fig. \ref{F-9}(d). The structures in Fig.~\ref{F-9}(e) show H2 moving between the two carbon centers at $t = 68.85$, 91.85, 94.45, with the the new \ce{C{2}-H{2}} bond formed at $t = 102.05$ fs. Owing to the excess kinetic energy remaining after the hop, the geometry continues to evolve for some time. By approximately $t = 300$~fs, the structural parameters (bond lengths and bond angles) stabilize and subsequently exhibit only small fluctuations. The structure at $t = 347.95$ fs shown in Fig. \ref{F-9}(e) represents a geometry similar to that of equilibrium acetaldehyde. \\

The entire trajectory is qualitatively consistent with the Gomer-Noyes mechanism \cite{GomerNoyes1950} and with previous simulations \cite{Tapavicza2008, Schmidt2020}. No further fragmentation is observed over the simulated interval. The newly formed \ce{C{2}-H{2}} bond remains close to 1.11 \AA, the \ce{C{1}-C{2}} bond remains intact, and the paired $\alpha$- and $\beta$-spin electronic structures persist to the end of the run. Acetaldehyde fragmentation into \ce{CH3^{.} + HCO^{.}}, reported by Gomer and Noyes \cite{GomerNoyes1950} and observed in the simulations of Tapavicza \textit{et al.} \cite{Tapavicza2008}, was not observed during our post-hop singlet trajectory. This behavior is consistent with that reported by Krenz \textit{et al.}, who observed subsequent \ce{C-C} cleavage to occur only in a small subset of the GS trajectories \cite{Schmidt2020}. \\

The two conditional post-hop pathways considered here lead to qualitatively different outcomes: the triplet trajectory undergoes direct \ce{C-C} fragmentation to \ce{H2CO + CH2}, whereas the open-shell singlet trajectory undergoes spin-symmetry restoration followed by intramolecular H transfer and formation of acetaldehyde.

\section{Conclusion}

We have investigated the ring-opening dynamics in oxirane and the subsequent product-forming reactions using TD$GW$-MD based on EQPT. The calculated ionization and low-lying excitation energies show good agreement with available spectroscopic data, supporting the electronic description used for the initial preparation. Following direct HOMO $\rightarrow$ LUMO excitation, assigned at the initial geometry to the low-lying $n\rightarrow3s$ (\ce{S1}) channel, the simulated trajectory undergoes asymmetric \ce{C-O} cleavage and ring-opening before the selected surface hop. This result demonstrates that direct preparation of the lowest excitation can initiate ring-opening within the present treatment, without requiring initial excitation into a higher-lying state. \\

Separate post-hop trajectories initiated from the ring-opened geometry reveal different reactions under the selected spin conditions. The broken-symmetry open-shell singlet trajectory forms acetaldehyde through intramolecular hydrogen transfer, whereas the triplet trajectory undergoes \ce{C-C} bond fragmentation to produce formaldehyde and methylene. The resulting product identities are consistent with those reported in oxirane photolysis experiments. These trajectories illustrate how the electronic configuration following ring-opening can lead to distinct rearrangement and fragmentation pathways. \\

The present study extends TD$GW$-MD to a reaction involving ring-opening, an open-shell intermediate, and subsequent changes in bonding. By examining both the initial excitation spectrum and the ensuing dynamics, it provides a many-body perspective on the reactivity following direct preparation of the lowest excitation in oxirane. Quantitative assessment of product branching will require statistical sampling together with evaluation of the probabilities of reaching the respective spin states. This assessment is left for future work.

\section{Acknowledgements}

This work used the Numerical Materials Simulator at the National Institute for Materials Science (NIMS) and computational resources of Fugaku provided by RIKEN through the HPCI System Research Project (Project IDs$\,$hp210218 and hp230468).
This work was supported by the grant-in-aid (KAKENHI) for Scientific Research B
(Grant Nos.$\,$23K21094 and 24K01149) from the Japan Society for the Promotion of Science (JSPS), AOARD (AFOSR) Grant No. FA2386-23-1-4061, and Innovative Science \& Technology Initiative for Security Project by Acquisition, Technology \& Logistic Agency (ATLA) grant number JPJ004596.

\section*{References}


\bibliographystyle{aipnum4-2-preservecase}
\bibliography{tdgw_bib}

\end{document}